\documentstyle[12pt]{article}
\newlength{\pubnumber} 

\catcode`\@=11
\@addtoreset{equation}{section}

\def\section{\@startsection{section}{1}{\z@}{3.5ex plus 1ex minus .2ex}
 {2.3ex plus .2ex}{\large\bf}}
\def\subsection{\@startsection{subsection}{2}{\z@}{2.3ex plus .2ex}
 {2.3ex plus .2ex}{\bf}}
\newcommand\Appendix[1]{\def\thesection{Appendix \Alph{section}}
 \section{\label{#1}}\def\thesection{\Alph{section}}}

\begin{document}

\begin{titlepage}
\samepage{
\setcounter{page}{1}
\rightline{IASSNS-HEP-95/25}
\rightline{UFIFT-HEP-95-26}
\rightline{NSF-ITP-95-130}
\rightline{\tt hep-th/9510223}
\rightline{October 1995}
\vfill
\begin{center}
 {\Large \bf String Unification,\\
   Higher-Level Gauge Symmetries,\\
   and Exotic Hypercharge Normalizations\\}
\vfill
\vspace{.25in}
 {\large Keith R. Dienes$^{1,3}$\footnote{
   E-mail address: dienes@sns.ias.edu},
   Alon E. Faraggi$^{1,2,3}$\footnote{
   E-mail address: faraggi@phys.ufl.edu},
   $\,$and$\,$ John March-Russell$^1$\footnote{
   E-mail address: jmr@sns.ias.edu}\\}
\vspace{.25in}
 {\it  $^{1}$ School of Natural Sciences, Institute for Advanced Study\\
  Olden Lane, Princeton, NJ 08540 USA\\}
\vspace{.05in}
 {\it  $^{2}$ Department of Physics, University of Florida\\
  Gainesville, FL  32611  USA\footnote{Current address.}\\}
\vspace{.05in}
 {\it  $^{3}$ Institute for Theoretical Physics, University of California\\
  Santa Barbara, CA 93106 USA\footnote{Temporary address.}\\}
\end{center}
\vfill
\begin{abstract}
  {\rm
  We explore the extent to which string theories with
 higher-level gauge symmetries and non-standard hypercharge normalizations
  can reconcile the discrepancy between
 the string unification scale and the GUT scale extrapolated from the
 Minimal Supersymmetric Standard Model (MSSM).
  We determine the phenomenologically allowed regions of $(k_Y,k_2,k_3)$
  parameter space, and investigate the proposal
    that there might exist string models with exotic
  hypercharge normalizations $k_Y$ which are less than
    their usual value $k_Y=5/3$.
  For a broad class of heterotic string
    models (encompassing most realistic string models which have been
    constructed), we prove that $k_Y\geq 5/3$.  Beyond this class,
    however, we show that there exist
    consistent MSSM embeddings which lead to $k_Y < 5/3$.
    We also consider the constraints imposed on $k_Y$ by demanding charge
    integrality of all unconfined string states, and show that only a limited
     set of hypercolor confining groups
    and corresponding values of $k_Y$
    are possible.  }
\end{abstract}
\vfill
\smallskip}
\end{titlepage}

\setcounter{footnote}{0}

\def\beq{\begin{equation}}
\def\eeq{\end{equation}}
\def\beqn{\begin{eqnarray}}
\def\eeqn{\end{eqnarray}}
\def\Tr{{\rm Tr}\,}
\def\KM{{Ka\v{c}-Moody}}

\def\ie{{\it i.e.}}
\def\etc{{\it etc}}
\def\eg{{\it e.g.}}
\def\half{{\textstyle{1\over 2}}}
\def\third{{\textstyle {1\over3}}}
\def\quarter{{\textstyle {1\over4}}}
\def\m{{\tt -}}
\def\p{{\tt +}}

\def\rep#1{{\bf {#1}}}
\def\slash#1{#1\hskip-6pt/\hskip6pt}
\def\slk{\slash{k}}
\def\GeV{\,{\rm GeV}}
\def\TeV{\,{\rm TeV}}
\def\y{\,{\rm y}}
\def\SM{Standard-Model }
\def\SUSY{supersymmetry }
\def\SSM{supersymmetric standard model}
\def\vev#1{\left\langle #1\right\rangle}
\def\l{\langle}
\def\r{\rangle}

\def\Htw{{\tilde H}}
\def\chibar{{\overline{\chi}}}
\def\qbar{{\overline{q}}}
\def\ibar{{\overline{\imath}}}
\def\jbar{{\overline{\jmath}}}
\def\Hbar{{\overline{H}}}
\def\Qbar{{\overline{Q}}}
\def\abar{{\overline{a}}}
\def\alphabar{{\overline{\alpha}}}
\def\betabar{{\overline{\beta}}}
\def\tautwo{{ \tau_2 }}
\def\calF{{\cal F}}
\def\calP{{\cal P}}
\def\calN{{\cal N}}
\def\smallmatrix#1#2#3#4{{ {{#1}~{#2}\choose{#3}~{#4}} }}
\def\bone{{\bf 1}}
\def\V{{\bf V}}
\def\b{{\bf b}}
\def\N{{\bf N}}
\def\bQ{{\bf Q}}
\def\t#1#2{{ \Theta\left\lbrack \matrix{ {#1}\cr {#2}\cr }\right\rbrack }}
\def\C#1#2{{ C\left\lbrack \matrix{ {#1}\cr {#2}\cr }\right\rbrack }}
\def\tp#1#2{{ \Theta'\left\lbrack \matrix{ {#1}\cr {#2}\cr }\right\rbrack }}
\def\tpp#1#2{{ \Theta''\left\lbrack \matrix{ {#1}\cr {#2}\cr }\right\rbrack }}


\def\inbar{\,\vrule height1.5ex width.4pt depth0pt}

\def\IC{\relax\hbox{$\inbar\kern-.3em{\rm C}$}}
\def\IQ{\relax\hbox{$\inbar\kern-.3em{\rm Q}$}}
\def\IR{\relax{\rm I\kern-.18em R}}
 \font\cmss=cmss10 \font\cmsss=cmss10 at 7pt
\def\IZ{\relax\ifmmode\mathchoice
 {\hbox{\cmss Z\kern-.4em Z}}{\hbox{\cmss Z\kern-.4em Z}}
 {\lower.9pt\hbox{\cmsss Z\kern-.4em Z}}
 {\lower1.2pt\hbox{\cmsss Z\kern-.4em Z}}\else{\cmss Z\kern-.4em Z}\fi}

\def\AEF{A.E. Faraggi}
\def\KRD{K.R. Dienes}
\def\JMR{J. March-Russell}
\def\NPB#1#2#3{{\it Nucl.\ Phys.}\/ {\bf B#1} (19#2) #3}
\def\PLB#1#2#3{{\it Phys.\ Lett.}\/ {\bf B#1} (19#2) #3}
\def\PRD#1#2#3{{\it Phys.\ Rev.}\/ {\bf D#1} (19#2) #3}
\def\PRL#1#2#3{{\it Phys.\ Rev.\ Lett.}\/ {\bf #1} (19#2) #3}
\def\PRT#1#2#3{{\it Phys.\ Rep.}\/ {\bf#1} (19#2) #3}
\def\MODA#1#2#3{{\it Mod.\ Phys.\ Lett.}\/ {\bf A#1} (19#2) #3}
\def\IJMP#1#2#3{{\it Int.\ J.\ Mod.\ Phys.}\/ {\bf A#1} (19#2) #3}
\def\nuvc#1#2#3{{\it Nuovo Cimento}\/ {\bf #1A} (#2) #3}
\def\etal{{\it et al,\/}\ }

\hyphenation{su-per-sym-met-ric non-su-per-sym-met-ric}
\hyphenation{space-time-super-sym-met-ric}
\hyphenation{mod-u-lar mod-u-lar--in-var-i-ant}


\setcounter{footnote}{0}
\section{Introduction and Summary}

String theory \cite{Sreviews}
is a unique candidate for the consistent unification of
quantum gravity with the gauge interactions.  Moreover, as a
fundamental theory of the
elementary interactions, string theory predicts a natural
unification of the corresponding couplings \cite{Ginsparg,scales}.
The scale at which such a unification takes place is related
to the Planck scale, and at the one-loop level
is found to be of the order \cite{vadim}:
\beq
       M_{\rm string}~\approx ~g_{\rm string}\,\times\,5\,\times\, 10^{17}
       {}~{\rm GeV}.
\label{eqthefirst}
\eeq
However, if one assumes that the particle spectrum
between the electroweak scale and the unification scale
is that of the Minimal Supersymmetric Standard Model (MSSM),
then the low-energy data predict a unification of the gauge
couplings at a scale which is of the order
\beq
       M_{\rm MSSM}~\approx ~2\,\times\,10^{16}~{\rm GeV}~.
\eeq
Thus, an order of magnitude separates the string unification scale
and the MSSM unification scale.
In other words, string-scale  unification
 --- together with the hypothesis that the spectrum below the string scale
is that of the MSSM ---
 predicts values for $\sin^2\theta_W(M_Z)$
and $\alpha_{\rm strong}(M_Z)$ which strongly disagree with the experimentally
observed values.
This is the well-known problem of string-scale gauge-coupling unification,
which is one of the more important issues facing string phenomenology.

It would seem that in an extrapolation of the gauge parameters over
fifteen orders of magnitude, a problem involving a single order of magnitude
would have many possible resolutions.  Indeed, in string theory there are
many possible
effects that can {\it a priori}\/ account for the discrepancy, and a
general
discussion of these effects can be found in Refs.~\cite{unif1,unif2}.
For example, there could
be large one-loop threshold corrections due to the infinite towers
of heavy string states which are otherwise neglected in an analysis of
the purely low-energy (massless) string spectrum.  Such threshold corrections
could redefine the string unification scale,
and potentially produce a new effective
scale in agreement
with the MSSM unification scale.  Alternatively, one could
contemplate that additional matter or gauge structure beyond the
MSSM could reconcile the two scales.
Another possibility, even within the MSSM gauge group and matter structure,
is that the boundary conditions for the gauge couplings at unification
could be altered by realizing the
$SU(2)$ and $SU(3)$ gauge factors
of the MSSM as affine Lie algebras with levels $k_2,k_3>1$.
A final possibility is that the normalization $k_Y$ of the
weak hypercharge current $U(1)_Y$ in string theory
is not
necessarily the one that is traditionally found in grand unified theories
(GUT's).  Hence, the potential exists for non-standard hypercharge
normalizations to alter the boundary conditions of the gauge couplings at
unification
in such a way as to resolve the discrepancy between the two unification
scales.

Despite these many possible approaches, however,
it turns out that the gauge coupling unification problem is
not simple to resolve within realistic string models.
While an analysis of the moduli dependence of the string threshold
corrections \cite{DKL} shows that
these corrections can indeed be substantial for large moduli,
a general expectation is that the string moduli settle near
the self-dual point, and are therefore naturally
small \cite{moduli}.
For small moduli, many numerical investigations
\cite{vadim,thresholdcalcs,unif2}
have shown that the string threshold
corrections are small, and thus do not
significantly affect the
string unification scale.
Indeed,
in Ref.~\cite{unif2}, general arguments were given that explain the suppression
of the string threshold correction in any modular-invariant string theory
without large moduli. Thus, without large moduli (or large products
of moduli\footnote{
   For $(2,0)$ string compactifications,
   it is effectively a {\it product}\/ of moduli that
   determines the size of threshold corrections \cite{modulitwozero}.
   It is then found \cite{NS}  that suitably chosen small values
   for individual moduli have the
   potential to yield larger threshold corrections.
   Unfortunately, this mechanism has not yet been realized
   within the context of realistic string models.}),
string threshold corrections are not likely to resolve the problem.

An alternative scenario is to contemplate the existence
of a simple GUT symmetry group between the MSSM unification scale and
the string scale.  However, despite several efforts \cite{stringguts},
no viable model employing this option has been constructed
to date.
Another possibility arises if there exists additional gauge structure beyond
the MSSM, arising at an intermediate energy scale.
Such additional gauge structure appears, for example,
in string models \cite{stringfsu5} realizing the flipped $SU(5)$
scenario \cite{fsu5}, or in string models \cite{ALR}
realizing the  Pati-Salam scenario \cite{patisalam}.
Such additional gauge structure may also appear in the form of extended
custodial
symmetries that are peculiar to certain string models \cite{custodial}.
Surprisingly, however, in all of these examples, a careful analysis shows
that the additional gauge structure at intermediate energy scales
only {\it increases}\/ the disagreement with the
experimental data \cite{unif1,unif2}.

Yet another possibility is that there exists additional
non-MSSM matter
in the desert between the electroweak
scale and the string unification scale \cite{GCU,Gaillard,BFY}.
In fact, the availability of such additional
thresholds in realistic free-fermionic models was demonstrated in
Ref.~\cite{GCU}.
Thus, in this respect, imposing the restriction that the spectrum below the
string scale is just that of the MSSM is {\it ad hoc}\/, and may be too
restrictive.
A detailed analysis of these matter states then shows
that while some of the realistic string models can achieve successful
string-scale unification if their non-MSSM matter states arise at appropriate
intermediate mass scales, in other models such matter cannot resolve
the discrepancy \cite{unif1,unif2}.

A final suggestion \cite{ibanez} is that in string theory,
the normalization of the weak hypercharge can generally differ
from that in grand-unified theories.
Indeed, if this hypercharge normalization $k_Y$
is carefully chosen relative to the non-abelian MSSM levels $(k_2,k_3)$,
the three gauge couplings will approximately unify at the
string scale rather than at the MSSM scale.  This is clearly a very
elegant and economical resolution of the problem.
Indeed, this scenario is as predictive as usual GUT unification,
since the introduction of the new string parameter $k_Y$
is offset by the new string-predicted relation (\ref{eqthefirst})
between the gauge coupling at unification and the unification scale.
However, it is found that in order for this scenario to
work within level-one string models,
this weak hypercharge normalization $k_Y$ should have the approximate
value $k_Y\approx 1.4$.
Unfortunately, despite various attempts \cite{otherefforts},
no models with $k_Y$ in this range have yet been constructed.
Furthermore, as we shall see, models with $k_Y$ in this range would
have immediate problems with fractionally charged states, and it is
therefore unclear whether such models would be realistic.

It is an important observation that hypercharge normalizations $k_Y\approx 1.4$
are smaller than the normalization factor $k_Y=5/3$ which appears
in simple grand-unified theories such as $SU(5)$ or $SO(10)$.
An important question, therefore, is
whether values $k_Y<5/3$ are realizable
in string models, or whether there
is some fundamental reason why they are not.  Of course,
from the point of view of the underlying
(rational) worldsheet conformal field theory,
$k_Y$ is indeed a free (rational) parameter,
and there are no further theoretical restrictions on the allowed values of
$k_Y$.
However, within a consistent string theory, other constraints can appear.

One such constraint arises due to the hypercharge of the positron.
In the Standard Model, the positron is a color- and electroweak-singlet.
Thus, the sole contribution to the conformal dimension of
the positron state from the Standard Model
gauge group factors is solely that from its weak hypercharge.  From this fact
and the fact that the positron must be realized as a massless string state,
it is possible to  show \cite{Schellekens} that any phenomenologically
consistent
string model must have $k_Y\geq 1$.

The question then arises as to whether other phenomenological constraints can
also be imposed on the value of $k_Y$.  A general expectation
might be that $k_Y=5/3$ is the minimal value allowed, simply because
the standard $SO(10)$ embedding is an
extremely economical way to embed all three generations with universal
weak hypercharge assignment.  Indeed, all of the realistic
string models constructed to date have $k_Y\geq 5/3$.
However, the possibility remains that there might
exist special, isolated string models
or constructions which manage to circumvent this bound, and populate the
range $1\leq k_Y<5/3$.

In this paper, we shall undertake a general
investigation of these issues.
Rather than look at specific string constructions,
we shall generally explore the extent to which string theories can
accommodate the higher-level gauge symmetries and/or
non-standard hypercharge normalizations that would be
necessary in order to reconcile the discrepancy between
string-scale unification and low-energy couplings.
Thus, we shall attempt a systematic analysis of the constraints
that govern which values of $(k_Y,k_2,k_3)$ are mutually realizable in
consistent realistic string models.

We begin by carefully analyzing the renormalization
group equations in order to determine the phenomenologically preferred
regions of $(k_Y,k_2,k_3)$ parameter space.  We find, as expected, that
values $k_Y/k_2 <5/3$ are preferred on the basis of gauge coupling
unification,
and that given the experimentally observed low-energy couplings,
only a narrow band of values for $(k_Y/k_2,k_3/k_2)$ are able
to achieve string-scale unification without large corrections
from other sources.
However, we also find a surprisingly strong correlation between
these {\it ratios}\/ of the levels $(k_Y,k_2,k_3)$, and
their absolute sizes.  For example, we find that
choosing the ratio $k_Y/k_2\approx 1.45-1.5$ is roughly consistent with
relatively small absolute sizes for the levels $(k_2,k_3)$,
but choosing $k_Y/k_2\approx 1.4$ requires levels $(k_2,k_3)$
which are significantly greater.  These results should therefore
provide strong constraints generally applicable to
realistic string model-building
employing higher-level gauge symmetries and/or
non-standard hypercharge normalizations.

Given these general results, we then focus on the
values that $k_Y/k_2$ may take in
realistic string models.
We follow essentially two ``orthogonal'' lines
of approach.

Our first approach
towards studying the possible values of $k_Y$ that may arise
for a given hypercharge group factor
involves analyzing the allowed embeddings of that hypercharge into a
consistent charge lattice.
Specifically,
by imposing the phenomenological requirement that the
entire MSSM spectrum appear and have the correct
hypercharge assignments,
we can examine the possible embeddings of the weak hypercharge in terms
of fundamental worldsheet currents.
In this way we are able to provide
various constraints on the value of $k_Y$.
We prove, for example,
that for a large class of realistic level-one string models
employing the so-called ``minimal'' MSSM embeddings,
one must have $k_Y\geq 5/3$.
This class includes most realistic free-field
string models which have been constructed in the literature.
However, it is nevertheless possible to extend such
an analysis beyond embeddings in this class, and we
find that we can construct consistent hypercharge embeddings
which manage to have $k_Y/k_2<5/3$
by using higher-twist sectors and/or higher levels $(k_2,k_3)$
for the non-abelian $SU(2)\times SU(3)$ MSSM gauge factors.
The special hypercharge embeddings that we construct
can therefore serve as a
useful starting point
in the construction of realistic $k_Y/k_2<5/3$ string models.

Our second approach is complementary to the first,
and instead examines the role that modular invariance plays
in restricting the mutually realizable values of $(k_Y,k_2,k_3)$.
As is well-known, string models typically contain a plethora
of unwanted massless chiral states carrying fractional electric charges,
and it is important for the purposes of realistic string model-building
to be able either to avoid such states altogether,
or to arrange to have them suitably confined under the influence
of an additional ``hypercolor'' gauge interaction.  However, it
turns out \cite{Schellekens}
that the possible scenarios by which this can be done are, through
modular invariance, highly dependent on the values of $(k_Y,k_2,k_3)$.
It is therefore
important to {\it correlate}\/ the allowed values of $(k_Y,k_2,k_3)$
with the appearance of fractionally charged states in the string
spectrum, and with the extra gauge interactions
under which such states might confine.
For example, it is straightforward to show that any level-one
$SU(3)\times SU(2)\times U(1)$ string model
 {\it without}\/ fractionally charged states must have $k_Y > 5/3$.
In this paper we shall analyze the more general situation
in which fractionally charged states can appear and are confined.
In this way we shall determine for which classes of fractionally
charged states and confining groups
the phenomenologically preferred values of $k_Y/k_2 < 5/3$
can be realized.

This paper is organized as follows.
In Sect.~2, we provide a self-contained review of
the definition of $k_Y$ in string theory, and the methods by
which it is calculated.  In Sect.~3, we then analyze the renormalization
group equations in order to determine the phenomenologically preferred
regions in $(k_Y,k_2,k_3)$ parameter space.
We then proceed, in Sects.~4 and 5, to investigate
the possible hypercharge embeddings that can be realized in
string theory.
Sect.~4, in particular, contains our proof that $k_Y\geq 5/3$ within a broad
class of realistic string models employing the ``minimal'' hypercharge
embeddings, and in Sect.~5 we go beyond this class in order to construct
consistent hypercharge embeddings with $k_Y/k_2 < 5/3$.
We then  turn, in Sect.~6, to the constraints imposed on $(k_Y,k_2,k_3)$ by
demanding the charge integrality of unconfined string states,
and in Sect.~7 we conclude with a summary of our results and
comments regarding various extensions.
Two background discussions have also been collected in Appendices.
Appendix A contains a review of the standard hypercharge
assignments for the MSSM states, and Appendix B contains the derivation
of a result quoted in Sect.~3 concerning the level-dependence of
two-loop corrections to the renormalization group equations.


\setcounter{footnote}{0}
\section{Technical Background:  Definition of $k_Y$}

Despite its seemingly simple role as the hypercharge normalization,
the factor $k_Y$ in string theory turns out to have an unexpectedly subtle
definition and interpretation.
In this section, therefore, we shall give a technical review of
this issue, and establish the normalizations that we shall
use throughout this paper.

The normalization factor $k_Y$ for the hypercharge group factor $U(1)_Y$
is similar in many ways to the so-called ``levels'' that appear
in the more general case
of non-abelian untwisted affine
Lie algebras.
We begin, therefore, by reviewing some basic facts
concerning these algebras, also known as \KM\
algebras \cite{KMalgebras}.
\KM\ algebras $\hat G$ are infinite-dimensional
extensions of the ordinary Lie algebras $G$, and contain the ordinary
Lie algebras as subalgebras.  They are generated by chiral worldsheet
currents $J^a(z)$ of conformal dimension $(1,0)$ satisfying the operator
product expansions (OPE's)
\beq
       J^a(z) J^b(w) ~=~ {{\tilde k}\,\delta_{ab}\over (z-w)^2}
     ~+~ {i\,f^{abc}\over z-w}\,J^c(w)~+~ {\rm regular}
\label{OPE}
\eeq
where $f^{abc}$ are the structure constants of the Lie algebra $G$
[with $a,b,c=1,...,{\rm dim}(G)$],
and where the first term on the right side (the ``central extension''
or Schwinger term with coefficient $\tilde k$) is the
new feature characterizing the \KM\ algebra.
It is clear that the special case with $\tilde k=0$
corresponds to the ordinary Lie current algebra.

For a non-abelian group ({\it i.e.}, one with non-vanishing
structure constants $f^{abc}$), one can define a unique normalization
for the currents $J^a(z)$ by fixing a particular normalization for the
structure constants and then demanding that the currents satisfy relations
of the form (\ref{OPE}).
One typically specifies the normalization of the
structure constants via
\beq
   \sum_{ab}\,f^{abc}\,f^{abd}~=~ C^{\rm (adj)}_{G} \,\delta^{cd}
\label{quadcas}
\eeq
where $C^{\rm (adj)}_G$ is the eigenvalue of the quadratic Casimir
acting on the adjoint representation.
This then fixes the normalizations of the currents, and by extension
fixes the value of the central extension coefficient $\tilde k$ as well
as the lengths of the root vectors $\lbrace \vec \alpha\rbrace$ of
the corresponding Lie algebra.
Alternatively, one can define the normalization-independent
quantities
\beq
    \tilde h_G ~\equiv~ {C^{\rm (adj)}_G \over \vec{\alpha_{h}}^2} ~,~~~~~
    k_G ~\equiv~ {2\,\tilde k \over \vec{\alpha_{h}}^2}
\label{coxeter}
\eeq
where $\vec \alpha_h$ is the longest root.
Here $\tilde h_G$ is the ``dual Coxeter number'' of
the group $G$,
and $k_G$ is the ``level'' of the \KM\ algebra.
We see, then, that the level $k_G$ of a \KM\ algebra has invariant
(intrinsic) meaning
only for a {\it non-abelian}\/ group $G$.
The central charge of the corresponding conformal theory
can then be similarly expressed:
\beq
        c_G ~=~ {k\, {\rm dim}(G)\over k+\tilde h_G}~,
\label{centralcharges}
\eeq
and likewise the conformal dimension of any given representation
of that algebra is given by:
\beq
           h_{(R)} ~=~ {C^{(R)}_G /\vec \alpha_h^2 \over k+\tilde h_G}
\label{confdim}
\eeq
where $C^{(R)}_G$ is the eigenvalue of the quadratic Casimir
acting on the representation $R$.
This eigenvalue is defined analogously to (\ref{quadcas}):
\beq
       \sum_{a=1}^{{\rm dim}(G)}\, (T^a T^a)_{ij}
            ~=~ C^{(R)}_G\,\delta^{ij}
\label{quadcasR}
\eeq
where $T^a$ are the group generators corresponding to
the representation $R$.

Given these results for non-abelian groups,
we now turn to the ``Ka\v{c}-Moody levels'' $k_Y$ associated with
abelian $U(1)$ groups.
In the case of the weak hypercharge $U(1)_Y$,
the corresponding value of $k_Y$ plays a crucial role in string theory
because, like the values of $k_3$ and $k_2$ corresponding to the
color and electroweak groups $SU(3)$ and $SU(2)$,
this value affects the
boundary conditions
of the gauge couplings at the string scale,
and consequently affects the predicted values of low-energy parameters such
as $\sin^2\theta_W$ and $\alpha_{\rm strong}$.  As we shall see, however,
the situation is somewhat more complicated for such
abelian group factors.

Strictly speaking, it is impossible to define
the intrinsic normalization-independent ``level'' $k$ at which
an abelian group is realized.  The difficulty arises as follows.
As we have discussed, the \KM\ level of
a non-abelian group is uniquely defined
as $k\equiv 2\tilde k/\vec \alpha_h^2$ where $\vec\alpha_h$ is
the longest root,
and where $\tilde k$ is the coefficient appearing in
the OPE's of its currents, as in (\ref{OPE}).
In the case of an {\it abelian}\/ factor, however, the
structure constants $f^{abc}$ all vanish, and consequently we
would find the corresponding OPE
\beq
   J(z) J(w)~~{{\buildrel {?}\over {=}}} ~~ {k\,\vec\alpha_h^2/2 \over
(z-w)^2}~+~
       {\rm regular}~.
\label{JJ}
\eeq
However, an abelian group factor contains no non-zero roots,
and therefore there is no longest root $\vec \alpha_h$ which can
be used to set a scale for this double-pole term or to absorb
changes in the normalization of the $U(1)$ current $J$.
Stated equivalently, we are prevented from fixing a normalization
for the currents by the absence of structure constants in
the algebra.
We see, therefore, that there is no invariant way of
defining the intrinsic \KM\ level of an abelian gauge
group factor.  Consequently, unlike the levels of
non-abelian group factors, the definition of
$k_Y$ (and with it the normalization of the corresponding
$U(1)$ current $J$) becomes a matter of {\it convention}.

The convention which is typically chosen makes use of the
fact that we are realizing such \KM\ algebras
in the particular physical context of string theory.
For {\it non}\/-abelian currents $J^a$, the normalizations are conventionally
fixed through the gauge couplings of these currents to each other
(through the three and four gauge boson vertices), and through their couplings
to
gravity (through the gauge-gauge-graviton vertex).
The corresponding \KM\ level $k$ then turns out to be the ratio between
these couplings \cite{Ginsparg}.
Indeed, this relative factor of $k_i$ arises because
the three gauge boson contributions arise from the single-pole
terms in the OPE (\ref{OPE}), while the gauge-graviton contributions
arise from the double pole.
We thereby obtain a relation of the form \cite{Ginsparg}
\beq
                8\pi {G_N\over \alpha'} ~=~ k_i\, g_i^2
\label{couplingratio}
\eeq
between the gravitational (Newton) constant
$G_N$ and the gauge coupling $g_i$ of any
non-abelian group factor.

For an {\it abelian}\/ current, however, we have no three gauge boson
couplings.   We nevertheless continue to have couplings
to gravity.  We therefore fix a proper normalization for a given $U(1)$
current $J$ by demanding that it have the same coupling
to gravity as any non-abelian current.
As shown in Ref.~\cite{Ginsparg}, this is tantamount
to requiring that $J$ take a normalization giving
rise to the OPE
\beq
   J(z) J(w)~=~ {1 \over (z-w)^2}~+~ {\rm regular}~.
\label{u1norm}
\eeq
An important point to notice here is that this is a {\it fixed}\/ normalization
which is independent of the lengths of the
roots of any of the non-abelian group factors.
Of course, we expect on physical grounds that this must be the case,
for it is possible to build string models for
which there are no non-abelian gauge group factors.
Such models will nevertheless exhibit couplings to gravity, however,
and therefore such a universal normalization for
the $U(1)$ currents can always be defined.

Given the normalization (\ref{u1norm}) for a proper $U(1)$ current,
the next issue is to define the ``level'' $k_1$ to which it corresponds.
As we have already discussed, there is no invariant way of defining the
level from the algebra itself [{\it e.g.}, we cannot use a relation
of the form (\ref{JJ})], and therefore we instead define the ``level'' $k_1$
based on physical terms
in analogy with the relation (\ref{couplingratio}).
In particular, given a $U(1)$ gauge group factor whose current is normalized
according to (\ref{u1norm}), we can calculate the ratio of the corresponding
gauge coupling $g_1$ to the gravitational coupling.  Such a calculation
yields
\beq
                8\pi {G_N\over \alpha'} ~=~ 2\,  g_1^2~.
\eeq
Therefore, by comparison with (\ref{couplingratio}), we define $k_J\equiv 2$
for any current
normalized according to (\ref{u1norm}).
Note that this then fixes a scale for the overall definition of the ``level''
$k_1$
corresponding to {\it any}\/ $U(1)$ current with {\it any}\/ arbitrary
normalization.
In particular, the factor $k_J$ corresponding to such an arbitrarily normalized
current
$J$ can be easily determined from the leading coefficient in its OPE expansion:
\beq
   J(z) J(w)~=~ {k_1/2 \over (z-w)^2}~+~ {\rm regular}~.
\label{k1def}
\eeq
We may therefore regard (\ref{k1def}) as  the general definition of
the ``level'' $k_1$ of any $U(1)$ current $J$.

Given this definition, it is then straightforward to determine the
levels $k_1$ and $k_G$ of each abelian or non-abelian gauge group factor
that arises in a given string model.
Recall that in a given heterotic string model, there are
typically 22 worldsheet currents which comprise the
the Cartan subalgebra of the total gauge group.
In a free-fermionic model-construction procedure, for example,
these elementary worldsheet Cartan currents $\hat J_i$ are in one-to-one
correspondence
with the internal worldsheet complex fermions $f_i$, with $\hat J\equiv
f_i^*f_i$
($i=1,...,22$).
Likewise, in a bosonic construction,
these Cartan currents are in one-to-one correspondence with the internal
worldsheet
bosons $\phi_i$, with $\hat J_i\equiv i\partial \phi_i$.
Now, in some string models, these Cartan currents $i\partial \phi_i$ may
combine with
other non-Cartan currents $\exp(i\alpha_i\phi_i)$ to fill out the adjoint
representation of a non-abelian gauge group;
in such cases the corresponding level $k_G$ can be directly computed via
(\ref{OPE}) and (\ref{coxeter}).
By contrast, there may also be various Cartan currents which do {\it not}\/
combine with any non-Cartan currents;  these then give rise to
elementary $U(1)$ factors in the gauge group.
However, since these Cartan currents $\hat J_i$ are always normalized
to satisfy (\ref{u1norm}), we see that any such elementary $U(1)$ gauge factors
will have $k_1=2$.

Similar considerations also hold for $U(1)$ gauge group factors whose currents
are realized as {\it linear combinations}\/ of the elementary Cartan
generators.
Indeed, in realistic string models, the gauge group factor $U(1)_Y$
corresponding to the Standard Model hypercharge will typically arise as
such a linear combination.  The corresponding values of $k_Y$ can nevertheless
be
determined in the manner discussed above.
In particular, let us suppose that the hypercharge current
$Y$ corresponding to the $U(1)_Y$ gauge factor is comprised of
the elementary worldsheet Cartan currents $\hat J_i$
via a linear combination of the form
\beq
        Y~=~ \sum_i \, a_i \,\hat J_i~
\label{genform}
\eeq
where the $a_i$ are certain model-specific coefficients
which describe the {\it embedding}\/ of the physical weak hypercharge current
in terms of the Cartan currents.
Of course, these coefficients $a_i$ and their overall normalization
must be chosen so that the eigenvalues of $Y$ will agree with the usual
hypercharge assignments for the MSSM states in the string
spectrum (see Appendix A).
However, since the elementary currents $\hat J_i$ are each
individually normalized so as to satisfy (\ref{u1norm}),
we see from (\ref{genform}) that the hypercharge
current $Y$ will have an OPE
of the form
\beq
   Y(z)\, Y(w)~=~ {\sum_i {a_i}^2 \over (z-w)^2}~+~ {\rm regular}~.
\eeq
Comparing with (\ref{k1def})
then allows us to identify the corresponding ``level'':
\beq
   k_Y~\equiv ~ 2 \,\sum_i \,{a_i}^2~.
\label{kYdef}
\eeq
Thus, in this way, we can define the level $k_Y$
corresponding to the hypercharge $U(1)_Y$
gauge group factor in a given string model.
Of course, as explained above, if $k_Y\not=2$ we must subsequently
renormalize $Y$ so that the properly normalized hypercharge
current
\beq
         \hat Y~\equiv ~\sqrt{2\over k_Y}\,Y
\label{rescaledY}
\eeq
satisfies the conventional OPE (\ref{u1norm}) with $k_{\hat Y}=2$.
In any case, it is clear from (\ref{kYdef}) that
\beq
         k_{cY}~=~ c^2 \,k_Y~
\label{krY}
\eeq
for any rescaling factor $c$, and thus such
rescalings are straightforward.

It is also straightforward to determine the conformal dimension
of a state with a given hypercharge.
For such an abelian group factor $U(1)_Y$,
the conformal dimension formula (\ref{confdim}) directly
reduces to
\beq
          h~=~ {Y^2\over k_Y}~.
\label{confdimY}
\eeq
However, it is important to verify by some independent means
that the result (\ref{confdim}) holds even for an abelian group.
This is most easily done by starting from the usual result
$h=Q^2/2$ for properly normalized $U(1)$ charges such as $\hat Y$,
and then rescaling $\hat Y$ as in (\ref{rescaledY}) to yield the result
(\ref{confdimY}) for $Y$.
As required,
this result for the conformal dimension is
invariant under rescalings
of $Y$ as a consequence of (\ref{krY}).

Before concluding, we emphasize once again
that this procedure for identifying the ``level'' $k_Y$
of the $U(1)_Y$ hypercharge gauge group factor
is ultimately a matter of convention, for
there is no intrinsic method of
defining the \KM\ level in the case of an abelian
group factor.  Consequently,
unlike the case of non-abelian
group factors, {\it we are not able to uniquely determine the
value of $k_Y$ without reference to the string model in
which this group factor is ultimately realized}\/.
Indeed, $k_Y$ is not an intrinsic function of the spacetime
gauge symmetries and associated algebras, but instead depends on
the {\it embedding}\/ of these symmetries within a
consistent string model.
Thus, the determination of $k_Y$ is highly model-dependent,
and in principle any rational value of $k_Y$
may be obtained.
There are, however, certain immediate bounds that can be placed on
the value of $k_Y$.  For example,
given the MSSM hypercharge assignments
listed in Appendix A,
we see that applying (\ref{confdimY}) to the $Y=1$
positron singlet state and requiring that this
state be massless (so that $h\leq 1$) trivially shows that we must
have $k_Y\geq 1$ in any string model containing such a massless state.
One of the main goals of this paper is to analyze what other constraints
may be placed on the possible values of $k_Y$
for realistic models containing the MSSM gauge group and spectrum.


\setcounter{footnote}{0}
\section{String Unification and Higher-Level
   Gauge Symmetries:  A Renormalization-Group Analysis}

We now explore the extent to which string-scale unification
might be reconciled with low-energy data by choosing
appropriate values for the three \KM\ levels
$(k_Y,k_2,k_3)$ which govern the MSSM
gauge group.  As we shall see, there are a variety
of constraints that come into play, and only a
limited number of possibilities are phenomenologically
viable.

As we have discussed in Sect.~2,
string theory predicts in general that at tree level,
the gauge couplings $g_i$ corresponding to each gauge group
factor $G_i$ realized at Ka\v{c}-Moody level $k_i$
will unify with the gravitational coupling constant $G_N$
according to \cite{Ginsparg}
\beq
     g^2_{\rm string}~=~8\pi \,{G_N\over\alpha'}~=~ k_i\, g_i^2
       ~~~~{\rm for~all}~i~
\label{treelevelrelation}
\eeq
where $\alpha'$ is the Regge slope.
This unification occurs at the Planck scale.
At the one-loop level, however, the string unification scale
is shifted down to \cite{vadim}
\beq
     M_{\rm string}~\approx ~g_{\rm string}\,\times\, 5\,\times \,10^{17} ~{\rm
         GeV}~,
\label{Mstring}
\eeq
and the relations (\ref{treelevelrelation}) are modified to
\beq
     {{16\pi^2}\over{g_i^2(\mu)}}~=~k_i\,{{16\pi^2}\over{g_{\rm string}^2}}~+~
     b_i\,\ln{ M^2_{\rm string}\over\mu^2}~+~\Delta_i^{\rm (total)}
\label{onelooprunning}
\eeq
where $b_i$ are the one-loop beta-function coefficients and
where the numerical value quoted in (\ref{Mstring}) is computed in
the $\overline{DR}$ renormalization-group scheme.
In (\ref{onelooprunning}), the quantities $\Delta_i^{\rm (total)}$
represent the combined corrections from various string-theoretic
and field-theoretic effects
such as
heavy string threshold corrections, light SUSY thresholds, intermediate
gauge structure, and extra string-predicted matter beyond the
MSSM.
Thus, in a given realistic string model,
we can use (\ref{onelooprunning})
to find the expected values of the
strong and electroweak gauge couplings at the $Z$-scale,
and thereby obtain explicit expressions for $\alpha_{\rm strong}(M_Z)$ and
$\sin^2\theta_W(M_Z)$.
Conversely, imposing the correct values of the low-energy couplings
at the low energy scale, we can determine for which values
of $(k_Y,k_2,k_3)$ we obtain a consistent string-scale unification.

For the present analysis we shall ignore
the contributions to the corrections $\Delta_i^{\rm (total)}$ which
arise from such sources as heavy string thresholds, light SUSY thresholds,
intermediate gauge structure, and extra non-MSSM matter.  Of course, such
corrections can be non-zero, and make contributions
to the running of the couplings.
Unfortunately, however, these corrections are also highly
model-dependent,
and therefore one cannot include them in such an analysis without
making recourse to a particular string
model.
Indeed, within the framework of a large
class of realistic string models,
such a complete analysis has been performed in Refs.~\cite{unif1,unif2},
and while most of these corrections were found to be small
(or, in general, insufficient to resolve the discrepancy between
the MSSM unification scale and the string scale), one of these
effects --- namely that arising from string-predicted
intermediate-scale matter beyond the MSSM ---
was found to be potentially large enough in certain models to resolve
the discrepancies.
Our present goal, however, is to determine
the extent to which string-scale unification is consistent in a general setting
 {\it without}\/ making recourse to such large corrections, but rather by
judiciously
choosing the \KM\ levels $(k_Y,k_2,k_3)$.
In other words, our purpose here is to
see to what extent a suitable foundation for string-scale unification
can be established {\it before}\/ such corrections are added.

We shall therefore
disregard these correction terms,
and thereby assume only the MSSM spectrum
between the $Z$ scale and MSSM scale.
We can then solve the equations (\ref{onelooprunning})
simultaneously in order to remove the direct dependence on $g_{\rm string}$.
In this way, we find that the low-energy couplings
$\sin^2\theta_W(M_Z)$ and $\alpha_{\rm strong}(M_Z)$
depend on the \KM\ levels $(k_3,k_2,k_Y)$
as follows:
\beqn
   \sin^2\theta_W(M_Z)  &=&
    {1\over {1+r}}\, \left\lbrack
      1~-~ \left(b_Y - {r}\, b_2\right)\,{a\over 2\pi}\,
      \ln {M_{\rm string}\over M_Z}
       \right\rbrack \nonumber\\
     \alpha_{\rm strong}^{-1}(M_Z) &=&
     {r'\over 1+r}\,\left\lbrack
    {1\over a} ~-~ \left( b_Y + b_2 - {1+r\over r'}\,b_3\right)
    \,{1\over 2\pi}\,
      \,\ln {M_{\rm string}\over M_Z}\right\rbrack~
\label{RGEs}
\eeqn
where $a\equiv \alpha_{\rm e.m.}(M_Z)$ is the electromagnetic coupling
at the $Z$ scale, and where
\beq
           r ~\equiv~ {k_Y\over k_2}
               ~~~~~~~ {\rm and}     ~~~~~~~
           r' ~\equiv~ {k_3\over k_2}~.
\eeq
Thus we see that
the magnitude of $\sin^2\theta_W$ depends on only the single
ratio $r$,
while the magnitude of $\alpha_{\rm strong}$ depends
on only the single ratio $r'/(1+r)$.
These observations imply that if we ignore the above corrections,
the value of $k_Y/k_2$ is can be determined purely by the value of
the coupling $\sin^2\theta_W$ at the $Z$ scale, while
the value of $k_3/k_2$ can then be adjusted in order to maintain
an acceptable value for $\alpha_{\rm strong}$ at the $Z$ scale.

It is straightforward to determine the numerical constraints
on these ratios.  For the MSSM spectrum (three generations and
two Higgs representations), we have
$b_Y=11$, $b_2=1$, and $b_3= -3$;  likewise the
experimentally measured values for the couplings \cite{couplings}
at the $Z$ scale are $\alpha_{\rm strong}(M_Z)|_{\overline{MS}}=0.120 \pm
0.010$
and $\sin^2\theta_W(M_Z)|_{\overline{MS}}=0.2315\pm 0.001$.
Note, however, that these values are determined in
the $\overline{MS}$ renormalization-group scheme,
whereas the value for $M_{\rm string}$ in (\ref{Mstring}) is given
in the $\overline{DR}$ scheme.
This means that although we are
ignoring the possible effects from heavy string threshold corrections,
light SUSY thresholds, intermediate gauge structure, and extra
non-MSSM matter when we set $\Delta_i^{\rm (other)}=0$ in
(\ref{onelooprunning}),
we should nevertheless retain the additional contributions to these quantities
which come from renormalization-group scheme conversion.
Likewise, there are also corrections from other model-independent sources
such as two-loop effects, and the effects of minimal Yukawa couplings.
These should also be included.
We shall see, in fact, that these effects can be quite significant.

In order to estimate these effects, we shall proceed as follows.
Let us denote by $\Delta_Y$, $\Delta_2$, and $\Delta_3$ the combined
corrections
to (\ref{onelooprunning})
from each of these sources.
On the basis of previous calculations \cite{unif1,unif2}, we expect
that the two-loop corrections are significantly larger (by an order of
magnitude)
than those from the Yukawa couplings or scheme conversion, and we shall see
this
explicitly.
To calculate the contributions to the $\Delta_i$ which come from two-loop
effects,
we evolve the MSSM couplings between
the $Z$ scale and the string scale with both the one-loop and Yukawa-less
two-loop
RGE's, assuming the MSSM spectrum with $(k_Y,k_2,k_3)=(5/3,1,1)$.
We then take the difference in order to obtain the values of the
two-loop corrections $\Delta_{Y,2,3}$.  In this way, we find the numerical
results
\beq
          \Delta_Y^{(\rm 2-loop)} ~\approx~ 15.5~,~~~~~~
          \Delta_2^{(\rm 2-loop)} ~\approx~ 15.1~,~~~~~~
          \Delta_3^{(\rm 2-loop)} ~\approx~ 7.8~.
\label{twoloopcorrections}
\eeq
Moreover, as we shall demonstrate in Appendix B, these quantities
depend only logarithmically on the ratios of the \KM\ levels $k_i$.
We can therefore take these quantities to be fixed in our subsequent
analysis.
Likewise, to calculate the corrections from the Yukawa couplings, we
evolve the two-loop RGE's for the gauge couplings coupled
with the one-loop RGE's for the heaviest-generation Yukawa couplings,
assuming
$\lambda_t(M_Z)\approx1.1$, $\lambda_b(M_Z)\approx0.175$, and
$\lambda_\tau(M_Z)\approx0.1$.  We then subtract the two-loop non-coupled
result.
This yields the following Yukawa-coupling corrections:
\beq
          \Delta_Y^{(\rm Yuk)} ~\approx~ -3.9~,~~~~~~
          \Delta_2^{(\rm Yuk)} ~\approx~ -2.7~,~~~~~~
          \Delta_3^{(\rm Yuk)} ~\approx~ -1.8~,
\label{Yukawacorrections}
\eeq
and we similarly expect the dependence of these quantities on the \KM\
levels $k_i$ to be at most logarithmic.
Finally, we can easily calculate the corrections of scheme conversion
between the $\overline{DR}$ and $\overline{MS}$ schemes,
for these corrections are
given directly in terms of the quadratic Casimirs of the
different representations.
We thus have
\beq
          \Delta_Y^{(\rm conv)} ~=~ 0,~~~~~~
          \Delta_2^{(\rm conv)} ~=~ 2/3~,~~~~~~
          \Delta_3^{(\rm conv)} ~=~ 1~,
\label{conversioncorrections}
\eeq
and these values are indeed strictly independent of
the \KM\ levels $k_i$.
Thus, combining the results (\ref{twoloopcorrections}),
(\ref{Yukawacorrections}), and (\ref{conversioncorrections}),
we have the total corrections
\beq
          \Delta_Y ~\approx~ 11.6     ~,~~~~~~
          \Delta_2 ~\approx~ 13.0    ~,~~~~~~
          \Delta_3 ~\approx~  7.0      ~.
\label{totalcorrections}
\eeq
We then find, including these corrections $\Delta_i$ into
(\ref{onelooprunning}),
that the two equations in (\ref{RGEs}) are likewise corrected
by corresponding terms $\Delta^{\rm (sin)}$ and $\Delta^{(\alpha)}$
respectively added to their right sides, where
\beqn
     \Delta^{\rm (sin)} &=& -\,{1\over 1+r}\,
    {a\over 4\pi}\,(\Delta_Y-r\,\Delta_2)\nonumber\\
     \Delta^{(\alpha)} &=& -\,{r'\over 1+r}\,
     {1\over 4\pi}\,\left(\Delta_Y+\Delta_2- {1+r\over r'}\,\Delta_3\right)~.
\label{sinalphacorrections}
\eeqn
Thus the primary level-dependence of these corrections is
that given in (\ref{sinalphacorrections}).

\input epsf
\begin{figure}[htb]
\centerline{\epsfxsize 3.5 truein \epsfbox {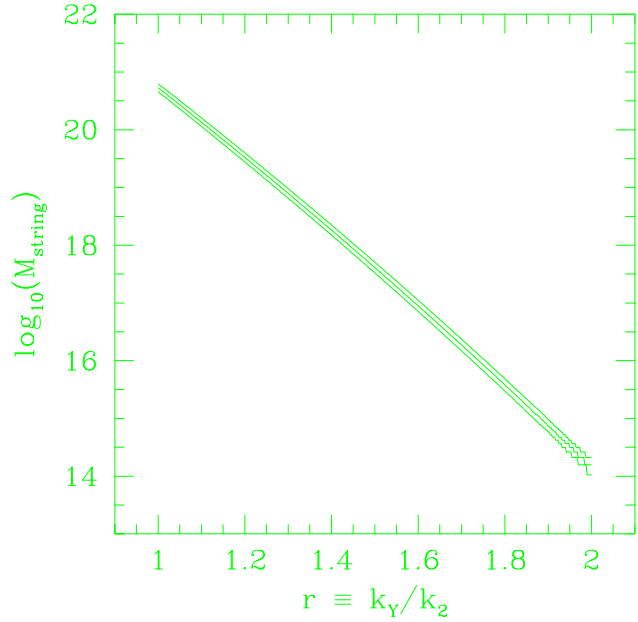}}
\caption{Dependence of unification scale $M_{\rm string}$ on
the chosen value of $r\equiv k_Y/k_2$.  The different curves
correspond to different values of $\sin^2\theta_W(M_Z)$, with
the lower curve arising for greater values.}
\label{rvslogM}
\end{figure}

\begin{figure}[htb]
\centerline{\epsfxsize 3.5 truein \epsfbox {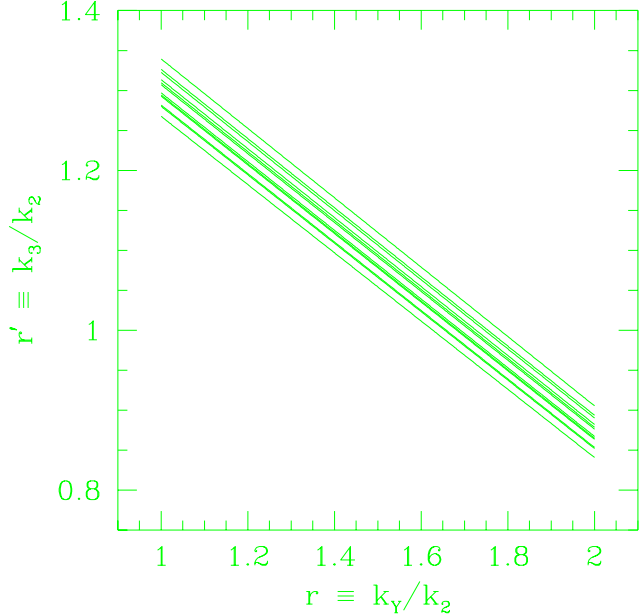}}
\caption{Values of $(r\equiv k_Y/k_2,r'\equiv k_3/k_2)$ yielding
the experimentally observed low-energy couplings.
The different curves correspond to different values of the
couplings, with the higher curves arising for smaller values of
$\sin^2\theta_W(M_Z)$ and $\alpha_{\rm strong}(M_Z)$, and the
lower curves arising for greater values.
Different points on any single curve correspond
to different unification scales.}
\label{rvsrprime}
\end{figure}

With these corrections included in the equations (\ref{RGEs}),
we can now determine the experimental constraints on
the ratios $r$ and $r'$.
Our results are shown in Figs.~\ref{rvslogM} and \ref{rvsrprime}.
In Fig.~\ref{rvslogM}, we show the dependence of
the scale of unification on the choice of the ratio $r$;
as expected, we find that $r=5/3$ leads to a unification scale approximately at
$M_{\rm MSSM}\approx 2\times 10^{16}$ GeV, while unification
at the desired string scale $M_{\rm string}\approx 5\times 10^{17}$ GeV
occurs only for smaller values of $r$, typically $r\approx 1.5$.
Note that this curve relies on only the low-energy input from
$a\equiv \alpha_{\rm e.m.}(M_Z)$ and $\sin^2\theta_W(M_Z)$.
We have taken $a^{-1}\equiv 127.9$ as a fixed quantity, and have varied
$\sin^2\theta_W(M_Z)$ in the range $0.2305 \leq \sin^2\theta_W(M_Z)\leq
0.2325$.

In Fig.~\ref{rvsrprime}, by contrast, we see how the value
of $r'$ must then be correspondingly adjusted
in order to maintain an experimentally acceptable value for
$\alpha_{\rm strong}(M_Z)$.
Thus, Fig.~\ref{rvsrprime}
summarizes those combinations of $(r,r')$
which are consistent with the phenomenologically acceptable
values for each of the three low-energy couplings, with
$\sin^2\theta_W(M_Z)$ allowed to vary in the above range,
and with $\alpha_{\rm strong}$ allowed to vary in the range
$0.115 \leq \alpha_{\rm strong}\leq 0.13$.
It is clear from this figure that decreases in $r$ must generally
be accompanied by increases in $r'$ in order to obtain
acceptable low-energy couplings;
moreover, only the approximate region $1.5\leq r\leq 1.8$ is capable
of yielding $r'\approx 1$.
This is an important constraint, for the \KM\ levels $k_2$ and $k_3$
are restricted to be {\it integers},
and thus arbitrary non-integer values of $r'$ would generally be possible
only for extremely high levels $k_2,k_3\gg 1$.

The analysis thus far has only constrained the values of
the {\it ratios}\/ of the levels $(k_Y,k_2,k_3)$, for
the renormalization group equations (\ref{RGEs}) give us no constraints
concerning the {\it absolute sizes}\/ of the \KM\ levels.
Fortunately, however, there is one additional constraint which must be imposed
in order to reflect the intrinsically stringy nature of the unification.
In field theory, there are ordinarily two free parameters
associated with unification:  the value of the coupling at the
unification scale, and the unification scale itself.  In string theory,
by contrast, these two parameters are tied together via (\ref{Mstring}).
Thus, in string theory it is actually not sufficient to determine the
ratios $k_Y/k_2$ and $k_3/k_2$
by merely demanding that they agree with low-energy data.
Rather, we must also demand that if our low-energy couplings are run up
to the unification point in a manner corresponding to certain values
of the levels $(k_Y,k_2,k_3)$,
the value of the predicted coupling $g_{\rm string}$ at the unification scale
must
be in agreement with the value of that unification scale.
If this final constraint is not met, then we have not achieved
a truly ``stringy'' unification.

This final constraint may be imposed as follows.
Since $a\equiv \alpha_{\rm e.m.}(M_Z)$ is the fixed ``input'' parameter
of our analysis, it is most convenient to follow the running of $\alpha_{\rm
e.m.}$.
Let us for the moment ignore the two-loop corrections $\Delta_i$.
Given the relation $1/\alpha_Y+1/\alpha_2=1/\alpha_{\rm e.m.}$
at all energy scales, we then find using (\ref{onelooprunning}) that
\beq
        {1\over \alpha_{\rm e.m.}(\mu)}~=~
        {1\over \alpha_{\rm e.m.}(M_{\rm string})} ~+~
               {b_Y+b_2\over 4\pi} \,\ln\,{M_{\rm string}^2\over \mu^2}~.
\label{alphaemrunning}
\eeq
Note, however, that at the unification scale,
we have
\beq
        \alpha_{\rm string}(M_{\rm string}) ~=~ (k_Y+k_2)~\alpha_{\rm
e.m.}(M_{\rm string})~.
\label{unifcouplings}
\eeq
Thus, taking $\mu=M_Z$ in (\ref{alphaemrunning}) and substituting
(\ref{unifcouplings}),
we find
\beq
        {1\over \alpha_{\rm string}(M_{\rm string})}~=~ {1\over k_Y+k_2}\,
        \left\lbrack {1\over  a} ~-~ {b_Y+b_2\over 4\pi} \,\ln\,{M_{\rm
string}^2\over M_Z^2}\right\rbrack~,
\label{neededlater}
\eeq
so that imposing (\ref{Mstring}) and defining $\alpha_{\rm str}\equiv
\alpha_{\rm string}(M_{\rm string})$
allows us to obtain the following transcendental
equation for $\alpha_{\rm str}$:
\beq
  {1\over \alpha_{\rm str}} ~=~ {1\over k_Y+k_2}\,\left\lbrack {1\over a}
     ~-~ {b_Y+b_2\over 2\pi} \,\ln\,{\sqrt{4\pi\alpha_{\rm str}}\,
  (5.27\times 10^{17}\,{\rm GeV}) \over M_Z\,({\rm GeV}) }\right\rbrack~.
\label{transcendental}
\eeq
Note that unlike the RGE's in (\ref{RGEs}), this equation depends
on the {\it absolute}\/ sizes of the \KM\ levels $k_i$.
Moreover, if we now restore the corrections $\Delta_i$
to (\ref{onelooprunning}), we find that the right side of this transcendental
equation
ultimately accrues a net correction
\beq
        \Delta^{(\rm trans)}  ~=~ -\,{\Delta_Y+\Delta_2\over k_Y + k_2}~.
\label{twoloopcorrectiontrans}
\eeq
This also depends on the absolute sizes of the levels $k_Y$ and $k_2$.
Thus, assuming that the $\Delta_i$ are themselves essentially independent
of the levels $k_i$,
we can solve this corrected transcendental equation numerically for
$\alpha_{\rm str}$
as a function of $k_Y+k_2$.  Then, by combining these results with our previous
results restricting the allowed parameters with respect to low-energy
couplings,
we can actually fix not only the ratios $k_Y/k_2$ and $k_3/k_2$ of the \KM\
levels,
but also their absolute values.

\begin{figure}[htb]
\centerline{\epsfxsize 3.5 truein \epsfbox {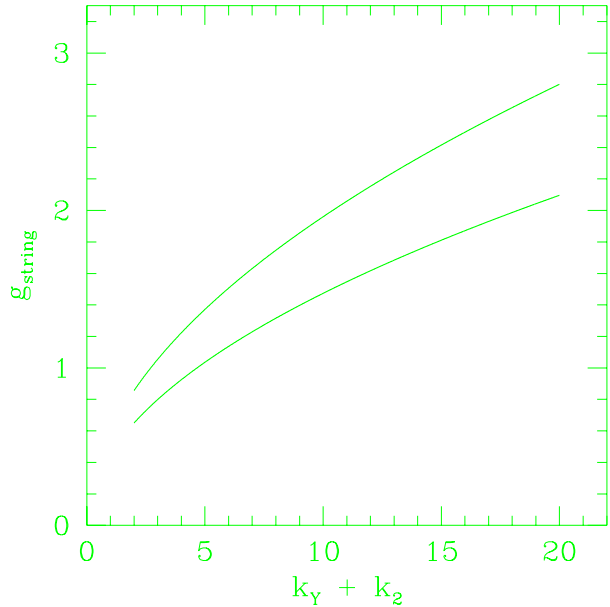}}
\caption{Solutions to the transcendental
equation (\protect\ref{transcendental})
for different values of $k_Y+k_2$, with the two-loop, Yukawa, and
scheme-conversion corrections included (upper curve),
and omitted (lower curve).}
\label{gstringplot}
\end{figure}

\begin{figure}[htb]
\centerline{\epsfxsize 3.5 truein \epsfbox {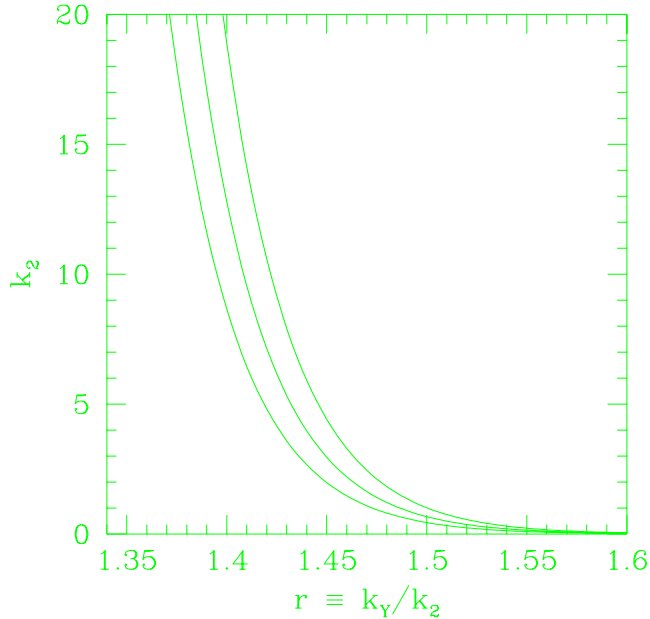}}
\caption{Dependence of $k_2$ on $r\equiv k_Y/k_2$.  The absolute
size of the \KM\ levels is set by the two-loop-corrected self-consistency
constraint
 (\protect\ref{transcendental}), in conjunction with the constraints from
the low-energy value of $\sin^2\theta_W(M_Z)$.
The different curves correspond to different values of
$\sin^2\theta_W(M_Z)$, with the lower/left curves arising for greater values.}
\label{rvsk2}
\end{figure}

Our results are shown in Figs.~\ref{gstringplot} and \ref{rvsk2}.
In Fig.~\ref{gstringplot}, we display the numerical solutions to
the transcendental self-consistency equation (\ref{transcendental}),
plotting $g_{\rm string}\equiv\sqrt{4\pi\alpha_{\rm str}}$
as a function of $k_Y+k_2$, both with and without
the two-loop, Yukawa, and scheme-conversion corrections.
As we can see, the effect of the corrections (\ref{twoloopcorrectiontrans})
is quite significant.
For $k_Y+k_2=5/3+1=8/3$, we find, as expected, that $0.7\leq g_{\rm string}
\leq 1.0$,
while for higher values of $k_Y+k_2$, the required string coupling increases
significantly.

Combining this result with that in Fig.~\ref{rvslogM},
we are finally able to correlate the acceptable values of
the level ratio $r\equiv k_Y/k_2$ with the absolute sizes of the levels.
Indeed, we see from Fig.~\ref{rvslogM} that the choice of a particular value of
$r\equiv k_Y/k_2$ not only sets a particular unification scale, but with that
choice of scale comes a certain scale for the string coupling via
(\ref{Mstring}),
and this in turn requires, via (\ref{transcendental}),
a certain absolute magnitude for the \KM\ levels themselves.
This combined dependence is shown in Fig.~\ref{rvsk2}, and is clearly quite
dramatic.
In particular, we see from this figure that values of $r$ in the range
$1.45\leq r\leq 1.5$
are consistent with relatively small \KM\ levels $k_{2,3}=1,2$, but
that for smaller values of $r'$, the required \KM\ levels increase
dramatically.
Interestingly, this figure also shows that there are natural
 {\it lower and upper bounds}\/ on phenomenologically viable values of $r$
if we wish to realize string-scale unification with only the MSSM spectrum.
The upper bound arises from the fact that $k_2\geq 1$,
which implies that $r\leq 1.5$, or
equivalently $k_Y\leq 1.5$.
The lower bound, by contrast,
arises if we wish to avoid unacceptably
large values of $g_{\rm string}$ at unification.
Specifically, Fig.~\ref{rvsk2} shows that if $r<1.35$, then
we are forced to realize this value of $r$ through extremely large
values of $k_2$ and $k_Y$, and from Fig.~\ref{gstringplot} we then see that
such large values of $k_2$ and $k_Y$ push the string coupling into
the non-perturbative regime.
Of course, we emphasize that the
precise placement of the curves in Fig.~\ref{rvsk2} depends quite strongly
on the exact value of $\sin^2\theta_W(M_Z)$ (as shown), as well as on
the various threshold corrections we have been neglecting.  Indeed, although we
expect these curves to remain essentially independent of these corrections
above $r\approx 1.45$, we see that the region $r<1.45$ is
 {\it exponentially}\/ dependent on the precise corrections.
Thus, these lower-$r$ regions of the curves in Fig.~\ref{rvsk2} are best
interpreted
as qualitative only.  Nevertheless, the general shapes of these curves,
evidently requiring extremely large changes in the absolute values of
$k_2$ and $k_3$ for seemingly modest shifts in the
ratio $r\equiv k_Y/k_2$,
are striking and should provide extremely strong constraints on
realistic string model-building.

Thus, combining all of these results,
it is clear that only certain tightly-constrained values
$(k_Y,k_2,k_3)$ are phenomenologically allowed if we are to achieve
string-scale unification with only the MSSM spectrum, and {\it without}\/
resorting
to large corrections from heavy string thresholds or
extra non-MSSM matter.  Indeed, from
Figs.~\ref{rvsrprime} and \ref{rvsk2},
we see that the best  phenomenologically allowed regions
of $(k_Y,k_2,k_3)$ exist in a narrow band stretching through
different values of $r$ in the range $1.4\leq r\leq 1.5$.
At the lower end of this band, for example, we have
values such as
\beq
        r\approx 1.4~~~~~\Longrightarrow~~~~~~
             (k_Y,k_2,k_3)~\approx~ (21,15,17)~,
\label{lowband}
\eeq
whereas at intermediate regions of the band
we have values such as
\beq
        r\approx 1.42~~~~~\Longrightarrow~~~~~~
             (k_Y,k_2,k_3)~\approx~ (14.2,10,11)~,
\label{midband}
\eeq
and at higher regions we have
\beq
        r\approx 1.5~~~~~\Longrightarrow~~~~~~
             (k_Y,k_2,k_3)~\approx~ (1.5,1,1)~{\rm or}~ (3,2,2)~.
\label{highband}
\eeq
It is interesting to note that, strictly speaking,
the {\it smaller}\/ values of
$r\approx 1.4$ are actually phenomenologically preferred, since they
require the larger absolute values of $(k_2,k_3)$ which in turn
enable us to more precisely achieve the required corresponding values of $r'$
shown in Fig.~\ref{rvsrprime}.
This is an important observation, for the allowed region indicated in
Fig.~\ref{rvsrprime}
encompasses two standard deviations in the low-energy couplings.
Thus, any failure to achieve the precise required value of $r'$
translates into a serious many-standard-deviation error in the
low-energy couplings.
Unfortunately, this low-$r$ region is particularly sensitive
to the corrections from, {\it e.g.}, heavy string threshold effects.
Thus, it is difficult to make reliable model-independent
statements concerning gauge-coupling unification in this region.

By contrast, for the purposes of {\it string model-building},
it turns out that smaller values for the non-abelian \KM\ levels,
such as $k_2,k_3=1 ~{\rm or}~2$, are strongly preferred on practical grounds.
Indeed, not only does it grow increasingly difficult to build string models
with higher values of $(k_2,k_3)$, but as these non-abelian levels
are increased, unwanted $SU(3)$ and $SU(2)$ representations
of increasing dimensionality begin to appear in the massless spectrum.
Thus, for the purposes of string model-building,
we find that the preferred regions
of parameter space are actually in the {\it higher}\/-$r$ region,
with $r\approx 1.5$, and with $k_2=k_3=1$ or $2$.
We would then hope that relatively {\it small}\/ corrections from other
sources such heavy string thresholds
would increase the effective value of $r'$ from $1$ to $r'\approx 1.05-1.1$.
Furthermore, in this higher-$r$ region, we do not expect such corrections to
have a major effect on the validity of the curves in Fig~\ref{rvsk2}.
We shall therefore focus our attention on
the values
\beq
       \cases{
        k_2~=~k_3~=~1 ~~{\rm or}~~ 2&~\cr
        r~\equiv~ k_Y/k_2~\approx~ 1.45 - 1.5~&~\cr}
\label{preferredvalues}
\eeq
in what follows.


\setcounter{footnote}{0}
\section{Hypercharge Embeddings}

We now turn to the central question that we seek to address
in this paper:  within the set of self-consistent ``realistic''
string models
containing the MSSM gauge group and particle representations
(as listed in Appendix A),
how might models be constructed with levels $(k_Y,k_2,k_3)$ in the
phenomenologically preferred region (\ref{preferredvalues})?
In particular, for realistic string models with levels $k_2=k_3=1$ or $2$,
what are the allowed values for the hypercharge normalization $k_Y$?
There are essentially two ``orthogonal'' approaches that we will follow.
In the next two sections, we shall focus on possible string embeddings
of the hypercharge group factor.
In this way, by imposing certain constraints,
we will arrive at a relatively small number of possible values of $k_Y<5/3$.
Indeed, one of the main results of this section will be a proof that within a
certain broad class of realistic level-one string models, one must
always have $k_Y\geq 5/3$.
In Sect.~6, by contrast, we shall instead focus on the implications of the
phenomenological requirement of charge integrality for string states.
As we shall see, this provides an alternative method of obtaining
general constraints on allowed values of $(k_Y,k_2,k_3)$.

\subsection{Hypercharge Embeddings:  General Strategy}

As discussed in
Sect.~2, the value of $k_Y$ in a given string
model is determined by the manner in which its hypercharge
group factor $U(1)_Y$ is ultimately embedded within the
given elementary $U(1)$ factors.
Therefore, in this section, we shall focus on the kinds
of hypercharge embeddings that are allowed in string
theory and that can potentially yield all of the
the MSSM representations with consistent hypercharge
assignments.

There are various strengths and weaknesses to this approach.
On the one hand, by focusing on stringy {\it embeddings}\/
rather than on any string {\it models}\/ themselves, our analysis
is in some sense independent of any particular string construction.
Thus, if we are able to prove that no suitable {\it embeddings}\/
with $k_Y<5/3$ exist (as we shall be able to do for a certain class
of string models), this result will therefore hold for all string
models in this class, whether constructed via free fermions,
free bosons, {\it etc}.
On the other hand, this approach does suffer from certain weaknesses.
Since an MSSM embedding is ultimately only a small part of a given string
model, an analysis of an MSSM embedding alone is often beyond
the reach of certain powerful {\it string}\/ consistency constraints
such as modular invariance.
Indeed, the string charge lattice
is generally of dimension 22, whereas the MSSM group is
typically embedded in only the first few dimensions of this lattice.
Without knowledge of what is happening in the remaining dimensions
of this lattice, we cannot impose the lattice constraints
(such as self-duality) that arise from modular invariance.
Therefore, if we are able to construct interesting MSSM embeddings
with $k_Y<5/3$ (as we will be able to do for certain other classes of
string theories),
the question will still remain as to whether there actually exist
string {\it models}\/ which realize these embeddings
in a manner consistent with modular invariance and other string
consistency conditions.

By contrast,
the alternative approach we shall follow in Sect.~6 will
incorporate modular invariance at an early stage.
Unfortunately, however, its power will be that of an existence proof:
it will be capable of determining under which conditions string
models with $k_Y<5/3$ might exist, but it will offer no clues
as to their constructions or embeddings.
Thus, we follow both approaches in the hope that some judicious
mixture of the two will be most fruitful in the long run.

Our analysis of potential hypercharge embeddings will proceed
in several stages.
First, we will examine the possible string embeddings of
the non-abelian factors of the MSSM gauge group,
namely $SU(3)_C$ and $SU(2)_L$.
This will ultimately allow us to identify the potential
string embeddings of the MSSM spectrum
({\it i.e.}, the realization of the MSSM representations as excitations
of underlying worldsheet fields),
which will in turn restrict their quantum numbers under the
constituent worldsheet $U(1)$ factors.
Given this information, we will then be able to examine to survey
possible classes of embeddings which enable a consistent hypercharge
assignment to be realized.   This will then enable us to deduce their
corresponding values of $k_Y$.

\subsection{Charge Lattices:  Basic Facts}

The most efficient method of discussing a particular string-theoretic
embedding is by specifying its associated {\it charge lattice}.
This may be defined as follows.  {\it A priori},
a heterotic string theory can give rise to gauge
groups of maximal rank $22$, and correspondingly there
exists a maximal set of $22$ Cartan generators $H_i$, $i=1,...,22$.
These generators are of course the elementary $U(1)$ currents we
discussed in Sect.~2;  for example, in  a free-fermionic model whose
internal sector is built out
of 22 complex worldsheet fermions $f_i$, these currents are
simply $H_i\equiv f^\ast_i f_i$.
Equivalent identifications may also be made for other constructions.
Indeed, in some cases (such as those involving necessarily real fermions),
the number of such Cartan generators surviving
the GSO constraints may actually be smaller than 22 (leading to the phenomenon
of {\it rank-cutting}, whereby the total rank of the resulting gauge group
is less than 22).
In any case, however, given a set of $r$ different Cartan generators $H_i$,
each state in the particle spectrum can then be described by specifying its
eigenvalues or {\it charges}\/ $Q_i$ with respect to these generators.
The set of such $r$-dimensional vectors $\bQ$ corresponding to all of
the states in the model then fills out the ``charge lattice''
of the model, and from this information the particle representations
and gauge group can be determined.
For example, the set of charge vectors corresponding to the gauge bosons
in the string spectrum fills out the root lattice of the corresponding
gauge group, and similarly the charge vectors corresponding to
the particles in a given multiplet fills out the weight system of the
corresponding
representation.
Since each direction of the charge lattice corresponds to one of the
underlying worldsheet $U(1)$ currents,
we therefore see that specifying the charge vector of a particular
state uniquely specifies its embedding, or equivalently its
realization in terms of
excitations of the underlying worldsheet fields.\footnote{
     Strictly speaking, this statement is true only for massless states;
     otherwise, one must specify both the charge vector and the
     energies of excitation.
     However, for our present purposes, we will only be considering the
      massless ({\it i.e.}, observable) states that comprise the MSSM
     spectrum.}

As we have already remarked,
working with the charge lattice of states is particularly convenient
because it provides a method of analysis which is largely independent
of the particular string model-building method of construction.
Indeed, the existence of a spacetime gauge group of rank $r$ implies the
existence
of such an $r$-dimensional charge lattice, and
the entire string spectrum must therefore fill out complete representations
with respect to this lattice.
Moreover, certain properties of this lattice can be deduced without
reference to the particular underlying string construction.
For example, modular invariance tightly constrains this lattice,
requiring that when it is tensored together with
a similar lattice from the supersymmetric side of the heterotic string,
the resulting product lattice must be {\it self-dual}\/.
Likewise, the requirement that the underlying
worldsheet conformal field theories be rational
implies that every charge vector must have {\it rational components}\/.
Furthermore, even the {\it lengths}\/ of the charge vectors are tightly
constrained.  For example, the squared lengths of the charge vectors
$\bQ$ corresponding to the non-Cartan gauge bosons of any simply-laced
group $G$ realized at level $k_G$
are constrained (in the normalizations conventional to string theory) to be
\beq
              \bQ \cdot \bQ ~=~ {2\over k_G}~,
\label{lengthlevel}
\eeq
and for non-simply laced groups this result holds for the gauge bosons
corresponding to the long roots.
As we shall see, all of these facts will prove crucial in our analysis.

\subsection{Gauge Group Embeddings}

We now turn to the possible string embeddings
of the non-abelian factors of the MSSM gauge group,
namely $SU(3)_C$ and $SU(2)_L$.
We shall assume, for the purposes of this analysis,
that these non-abelian group factors are realized
at \KM\ levels $k_2=k_3=1$.

We begin with the embedding of the $SU(2)$ factor.
As we have seen in Sect.~2, such a group
factor has central charge $c=1$, and likewise has rank $r=1$.
Now, in general each dimension of the charge lattice corresponds
to central charge $c=1$;  this is true for all models built from
worldsheet bosons (regardless of their method of compactification), or
worldsheet fermions (complex or real).
Indeed, this is true for all of the
free-field model-construction procedures through which phenomenologically
appealing realistic string models have been constructed.
Consequently, we see that an $SU(2)$ gauge group factor at level
$k_2=1$ must be realized through a lattice of at least dimensionality one.
However, we see from (\ref{lengthlevel}) that for level $k_2=1$, a
one-dimensional
lattice would require the non-Cartan $SU(2)$ gauge boson
to have a charge of the form $\bQ=(\sqrt{2})$, which fails to
contain rational components.
Thus, we see that $SU(2)$ at level one cannot be realized completely within a
single
dimension of the charge lattice, and must be realized instead as a non-trivial
embedding across several dimensions.  In this case, however, it is easy to
see that there {\it does}\/ exist a suitable {\it two-dimensional}\/ embedding,
with non-Cartan charges
\beq
            (-1,1) ~~~~{\rm and}~~~~~ (1,-1)~.
\eeq
[By changing the relative orientation of these dimensions, this is equivalent
to $(1,1)$ and $(-1,-1)$, but the above orientation is chosen for future
convenience.]
As required, these charge vectors have only rational components.
Thus, we see that the minimal embedding of the $SU(2)_1$ root lattice is one
in which the $SU(2)$ axis lies along the diagonal between
  {\it two}\/ dimensions of the total charge lattice.
The orthogonal direction $(1,1)$ can then correspond to an additional
$SU(2)$ factor, or a $U(1)$ factor.  We will denote the ``unit'' charge
vector in this orthogonal direction as $\bQ_L\equiv (1,1)$.

A similar analysis can be performed in order to determine the minimal
$SU(3)$ embedding.  Since $SU(3)_1$ has central
charge and rank equal to two, {\it a priori}\/ only
two dimensions of the charge lattice are needed.
However, as expected, no such embedding of the $SU(3)$ root system
with rational components exists, and instead three dimensions of the
charge lattice are required.  The three-dimensional embeddings
for the two simple roots  of the $SU(3)$ root system are
then
\beq
        (-1,1,0)~~~~~{\rm and}~~~~~~
        (0,-1,1)~,
\eeq
and it is easy to verify that these two vectors have a relative angle
of $120^\circ$, as required.
The remaining positive non-Cartan generator for $SU(3)_1$ is then
the sum of these roots, or $(-1,0,1)$.
As in the $SU(2)$ case, the direction perpendicular to the $SU(3)$
plane corresponds to an additional $U(1)$ factor, and we shall
choose a ``unit'' charge vector in this orthogonal direction
to be $\bQ_C\equiv (1,1,1)$.
Note that in this $SU(3)$ case,
the extra orthogonal direction cannot correspond to an $SU(2)$ factor, since
$\bQ_C$ does not lie diagonally between two
directions of the charge lattice [{\it i.e.}, it is not of the
form $(1,1,0)$ or permutations thereof].

These results imply, therefore, that the minimal embedding of the
gauge group $SU(3)\times SU(2)$ at levels $k_3=k_2=1$ in string
theory requires a {\it five-dimensional}\/ lattice.
We shall henceforth take these to be the first five directions
of the charge lattice, with eigenvalues $Q_1,Q_2,Q_3$ pertaining
to the $SU(3)$ embedding, and eigenvalues $Q_4,Q_5$ pertaining
to the $SU(2)$ embedding.
As discussed above, we then have $Q_C\equiv Q_1+Q_2+Q_3$ and
$Q_L\equiv Q_4+Q_5$.
We emphasize, however, that this particular embedding is only the
simplest (and most compact) embedding for the $SU(3)_C$ and $SU(2)_L$
group factors realized at levels $k_2=k_3=1$.  There are, of course,
many other potential embeddings for these gauge group factors which
involve many more than five components of the charge lattice.
This embedding, however, is ``minimal'' in that it involves the
fewest number of elementary dimensions of the charge lattice,
and is therefore most likely to yield the smallest values of $k_Y$.

\subsection{Embeddings for matter representations}

Given the above embeddings for the root systems ({\it i.e.}, for the adjoint
representations), it is now straightforward to determine
the possible embeddings for the matter representations that concern
us, namely the $\rep{2}$ of $SU(2)$, and the $\rep{3}$ and $\rep{\overline{3}}$
of $SU(3)$.
Since the lengths of the weight vectors of the doublet
representation of $SU(2)$ are half those for the adjoint,
we immediately see that the doublet of $SU(2)_1$ is realized
as:
\beq
        \rep{2} ~{\rm of}~ SU(2):~~~~~~~~ \pm (-1/2,+1/2)~.
\label{doubletembedding}
\eeq
Likewise, since the squared lengths of the weight vectors of the triplet
representation of $SU(3)$ are one third those for the adjoint,
we immediately see that given the above embedding for the
$SU(3)$ roots, we have the following embedding for the $SU(3)$
triplet:
\beq
       \rep{3}~{\rm of}~SU(3): ~~~~~~~~ \cases{ (-1/3,-1/3,+2/3)~&~ \cr
      (-1/3,+2/3,-1/3)~& ~\cr
      (+2/3,-1/3,-1/3)~.&~\cr}
\label{tripletembedding}
\eeq
The $\rep{\overline{3}}$ representation of
$SU(3)$ is of course given by the negatives
of these three weights.

In constructions of string models based on worldsheet
bosonic degrees of freedom,
the denominators of the components of the charge lattice
are related to the bosonic radii of compactification.
Likewise, in a free-fermionic construction of string models,
these denominators are related to the boundary conditions of the worldsheet
fermions as they traverse the non-contractible cycles of the torus.
Indeed, in constructions employing only periodic (Ramond) or anti-periodic
(Neveu-Schwarz) fermions, the greatest denominator that the charge lattice
can have is $2$.
This does not mean, however, that $SU(3)$ triplets cannot be obtained
in such models;  rather, this observation implies that if such $SU(3)$ triplets
do appear, they must also carry a non-zero quantum number under the
orthogonal $U(1)_C$ generator $Q_C\equiv(1,1,1)$.
For example, we see that the $SU(3)$ triplet
with $U(1)_C$ quantum number $1/3$ corresponds to charge vectors
of the form $(0,0,1)$ and permutations.
Such states with integer charge components are typically
obtained from Neveu-Schwarz sectors.
Similarly, the triplet with $U(1)_C$ quantum number
$-1/6$ will have charge vectors of the form $(-1/2,-1/2,+1/2)$,
which will typically be obtained in a Ramond sector.
However, we see from the general considerations above that
such non-zero $U(1)_C$ quantum numbers are {\it required}\/
if we are using only periodic or anti-periodic fermions and
the minimal $SU(3)$ embedding.

Thus, in general, we see that
we must allow our $SU(3)$ and $SU(2)$ representations to simultaneously
carry charges under $U(1)_C$ and $U(1)_L$ respectively.
Since these $U(1)$ gauge factors correspond to the $(1,1,1)$ and
$(1,1)$ directions respectively, we see that
if the $SU(2)$ doublet has $U(1)_L$ charge $q_L$
and the $SU(3)$ triplet has $U(1)_C$ charge $q_C$,
then these representations correspond to the charge vectors:
\beqn
        (\rep{2})_{q_L} ~{\rm of}~ SU(2):&~~~~~~~~ &\pm (-1/2,+1/2)~+~
q_L\,(1,1)~\nonumber\\
       (\rep{3})_{q_C}~{\rm of}~SU(3): &~~~~~~~~ &
           \left\lbrace \matrix{ (-1/3,-1/3,+2/3) \cr
                 (-1/3,+2/3,-1/3)\cr
                 (+2/3,-1/3,-1/3)\cr} \right\rbrace ~ +~ q_C\,(1,1,1)~.
\label{fundreps}
\eeqn
Likewise, since the uncharged singlet representations correspond to
$(0,0,0)$ and $(0,0)$, the corresponding charged representations correspond
to $q_C(1,1,1)$ and $q_L(1,1)$ directly:
\beqn
        (\rep{1})_{q_L} ~{\rm of}~ SU(2):&~~~~~~~~ & q_L\,(1,1)~\nonumber\\
       (\rep{1})_{q_C}~{\rm of}~SU(3): &~~~~~~~~ & q_C\,(1,1,1)~.
\label{singlets}
\eeqn

\subsection{Conformal Dimensions}

Given these representations of $SU(3)$ and $SU(2)$, our next step
is to examine their conformal dimensions.
This is relevant because massless string states must have total conformal
dimensions equal to $1$, and this in turn will constrain the allowed
values for the $U(1)_C$ and $U(1)_L$
quantum numbers that such representations can carry.
For string models without rank-cutting ({\it i.e.},
models with rank-22 simply-laced gauge groups),
the conformal dimension $h(\bQ)$ of a given state can in general
be determined
from its charge vector \bQ\ via the relationship
\beq
          h(\bQ) ~=~ {1\over 2}\, \bQ\cdot \bQ~.
\label{chargedotproduct}
\eeq
Indeed, this relationship holds for all states except those
adjoint gauge boson states in the Cartan subalgebra.
Thus, calculating the conformal dimensions of the representations in
(\ref{fundreps}) and (\ref{singlets}), we have
\beqn
  SU(3)\times U(1)_C:&&~~\cases{
         ({\rep{1}})_{q_C}~:&  ~~~~$h~=~ (3/2)\,q_C^2$\cr
         ({\rep{3}})_{q_C}~:&  ~~~~$h~=~ 1/3~+~ (3/2)\,q_C^2$\cr}\nonumber\\
  SU(2)\times U(1)_L:&&~~\cases{
         ({\rep{1}})_{q_L}~:&  ~~~~$h~=~ q_L^2$\cr
         ({\rep{2}})_{q_L}~:&  ~~~~$h~=~ 1/4~+~ q_L^2~$\cr}.
\label{confdims}
\eeqn
Note that these results could also have been obtained directly
from (\ref{confdim}).

There are certain immediate lessons that can be drawn from these results.
A trivial result, for example, is that there are natural bounds on the
values of $q_C$ and $q_L$ that each representation can carry while still
remaining massless.
A more profound result, however, is the observation that
in string theory, we cannot
realize all of our MSSM representations
as singlets under additional ({\it i.e.}, hidden) gauge group factors
in the string model.
Rather, we are {\it forced}\/ to have at least some of our MSSM representations
transform non-trivially under some other hidden factors beyond
$SU(3)\times SU(2)\times U(1)_C \times U(1)_L$.
This is most easily seen by examining the lepton $L\equiv(\rep{1},\rep{2})$
representation.
If this state were to be realized as a singlet under all non-MSSM gauge
factors,
then the contribution to its conformal dimension from the MSSM factors
would itself be equal to one, or
\beq
              {3\over 2}\, q_C^2 ~+~ q_L^2
           ~~{{\buildrel {?}\over {=}}}~~ {3\over 4}~.
\eeq
Remarkably, however, this equation has no solution for rational values of
$(q_C,q_L)$.
Hence, some of the total conformal dimension of this representation must be
attributed to string excitations beyond the degrees of freedom giving
rise to the MSSM factors.
In general, all of the MSSM representations will have some non-MSSM
conformal dimensions as well.
Hence the extra (hidden) gauge group factors play a non-trivial role in
realizing the MSSM states as massless excitations in string theory.

\subsection{Chirality}

We now discuss the implications of the fact that our representations
(\ref{representations}) are to be {\it chiral}\/.
This means, of course, that while we not only demand that the representations
listed in (\ref{representations}) actually appear, we must also simultaneously
demand that their complex conjugates {\it not}\/ appear.
In terms of the charge lattice of vectors $\bQ$, this chirality constraint
means the following.
In general, the full charge lattice of a given string theory is
a Lorentzian lattice of dimension $(r',r)$ where
the first $r' \leq 10$ dimensions correspond to the
right-movers (the supersymmetric side of the heterotic string),
and the remaining $r\leq 22$ dimensions correspond to the
left-movers (the internal bosonic side of the heterotic string).
The first of the right-moving charge components corresponds to the spacetime
statistics of a given state (integer for bosons, integer $+ 1/2$ for fermions),
and the first five left-moving components are simply the five-dimensional
charge vectors $\bQ$ that we have been discussing above.
Now, CPT invariance of the string spectrum implies that for every
$(r',r)$-dimensional
charge vector $\bQ$ that appears in the string spectrum, there must also appear
the charge vector $-\bQ$.
Chirality, by contrast, involves all but the first (spacetime) component of the
charge vector, and implies that for any given charge
vector $\bQ=(Q_1,Q_2,...Q_{r'+r})$, the chiral conjugate
$\tilde \bQ\equiv (+Q_1,-Q_2,...,-Q_{r'+r})$ must not appear.

In general, this is a difficult requirement to implement in a general fashion,
and whether a particular representation appears chirally or non-chirally
usually depends on the intricate details of the GSO projections in the
string model in question.
One general fact that we will use, however,
is that any string state whose internal charge vector $\bQ$
is of the form $(\pm 1,\pm 1,0,0,...,0)$ or permutations
cannot be chiral.  This observation follows from the fact that such
a left-moving charge vector is also the left-moving
charge vector of a gauge boson in the theory,
and such gauge bosons (which belong to the adjoint) necessarily are non-chiral.
This assertion holds whether this hypothetical gauge boson state
is actually in the string spectrum, or is GSO-projected out
({\it i.e.}, whether the gauge group to which it corresponds
is preserved, or broken).
In either case
such a state will either appear with its conjugate,
or not appear at all.
Thus, the chirality of our MSSM
representations demands that no MSSM matter state
can have the form
\beq
       \bQ_{\rm non-chiral}~=~(\pm 1,\pm 1,0,0,0,...0)~~~~{\rm
or~permutations}~.
\label{nonchiralQ}
\eeq

\subsection{Hypercharge Embeddings}

Given the above minimal embeddings for the $SU(2)$
and $SU(3)$ adjoint and fundamental representations, we can now
put the pieces together to examine the potential hypercharge embeddings
which successfully reproduce
the required hypercharge assignments.

As we have seen above,
in string theory we cannot assume that the MSSM representations are
charged under only the MSSM gauge group;  rather, we are forced
to have them transform non-trivially under
the additional ({\it i.e.}, hidden) gauge group factors
in the string model.
Thus, the MSSM representations will in general have non-zero quantum numbers
({\it i.e.}, non-zero charge components $Q_i$)
beyond the first five Cartan generators corresponding
to the $SU(3)\times SU(2)$ embedding.

Despite this fact, we will
restrict the present analysis to weak hypercharges $Y$
that take the ``minimal'' form
\beq
        Y~\equiv~ \sum_{i=1}^{5} \,a_i\,Q_i~.
\label{simpleform}
\eeq
This restriction implies, of course,
that $Y$ is independent of any extra hidden gauge group factors.
Note that this does not contradict our assertion that
the MSSM representations must be charged under such factors,
for it is possible that the {\it weak hypercharge}\/ does
not depend on these extra factors.
Indeed, many of the realistic string models which exist in
the literature have hypercharge embeddings of the form (\ref{simpleform}).
Furthermore, since this is in some sense the most ``compact''
distribution of the hypercharge --- involving the MSSM components
of the lattice only --- we may expect that this form is likely
to yield the smallest values of $k_Y$.  In any case,
since we have no model-independent information regarding the
remaining ``hidden'' components of the
charge lattice (other than self-duality
of the {\it entire}\/ 32-dimensional left/right charge lattice),
it is clear that we are unable to perform any detailed model-independent
analysis for hypercharge embeddings beyond
those of the minimal form (\ref{simpleform}).

Given the form (\ref{simpleform}), then,
our goal is to determine the possible solutions for the $a_i$
coefficients.  The corresponding value of $k_Y$ is then determined
via (\ref{kYdef}).

The first thing we notice is that since the value of $Y$ must be
the same for all members of a given MSSM representation,
$Y$ must have the same eigenvalue for all members of an $SU(3)$ triplet
or $SU(2)$ doublet.  This immediately implies that
\beq
      a_1=a_2 =a_3 ~~~~~{\rm and}~~~~~ a_4=a_5 ~.
\eeq
Let us denote $a_1=a_2=a_3\equiv A_1$ and $a_4=a_5\equiv A_2$.
Then acting on {\it any}\/ $SU(3)\times SU(2)$ representation,
we simply find
\beq
        Y \, (\rep{R}_{q_C}, \rep{R'}_{q_L} )
       ~=~ 3 A_1 q_C ~+~ 2 A_2 q_L~.
\label{YCL}
\eeq
Hence, only the $q_C$ and $q_L$ eigenvalues are relevant for $Y$.
In order to solve for $A_1$
and $A_2$, therefore, we must first determine the
$q_C$ and $q_L$ quantum numbers of each MSSM representation.

In general, there are many possibilities for the charges $(q_C,q_L)$
for each of the MSSM representations (\ref{representations}),
for these charges depend on the details of the underlying string
model.  Likewise, for each different combination,
there may or may not exist a solution $(A_1,A_2)$
which describes a corresponding successful hypercharge embedding.
Hence, we must first narrow down the list of $(q_C,q_L)$ possibilities.
There are a variety of means through which this can be done.
First, we recall of course that the rationality of the worldsheet
conformal field theory requires that these charges be rational numbers.
Second, we see from (\ref{confdims})
that for our MSSM states to be potentially massless, the
$(q_C,q_L)$ quantum numbers  of each $SU(3)\times SU(2)$ representation
must satisfy the constraints:
\beqn
   (\rep{3},\rep{2})~:&~~~~~~~~  (3/2)\,q_C^2 + q_L^2 &\leq ~5/12 \nonumber\\
   (\rep{\overline{3}},\rep{1})~:&~~~~~~~~  (3/2)\,q_C^2 + q_L^2 &\leq~ 2/3
\nonumber\\
   (\rep{1},\rep{2})~:&~~~~~~~~  (3/2)\,q_C^2 + q_L^2 &\leq~ 3/4 \nonumber\\
   (\rep{1},\rep{1})~:&~~~~~~~~  (3/2)\,q_C^2 + q_L^2 &\leq~ 1~.
\eeqn
Third, in order for the representation $(\rep{R}_{q_C},\rep{R'}_{q_L})$
to be potentially chiral,
the corresponding five-dimensional charge vector $\bQ=(Q_1,Q_2,Q_3,Q_4,Q_5)$
must not be of the form (\ref{nonchiralQ}).  Given the results in
(\ref{fundreps})
and (\ref{singlets}), this too rules out certain combinations of $(q_C,q_L)$.

Finally, we must impose a {\it moding requirement}\/ on the charge
vectors $\bQ$ corresponding to each representation.
In general, this means that given a certain moding for the
underlying worldsheet fields of the string,
there exists a certain corresponding value
for the greatest common denominator $\Delta$ of all of the elements $Q_i$
of the charge vectors of the corresponding spacetime states.
For example,
if a string model is built in the free-fermionic construction
using fermions with only
Ramond or Neveu-Schwarz (periodic or anti-periodic)
boundary conditions around the non-contractible cycles of the
torus, then we have $\Delta=2$, and all of the allowed states in such a model
must have charge
vectors whose components are integers or half-integers only.
In general $\Delta$ must be a positive even integer.

Since the analysis of all possible hypercharge configurations
is highly dependent on the particular value of $\Delta$ chosen, we must
examine each case in turn.
For each value of $\Delta$, we shall
therefore simply enumerate all possible values
of $(q_C,q_L)$ for each MSSM representation subject to the
masslessness, chirality, and moding constraints,
and for each set of combinations of $(q_C,q_L)$ we will then
determine whether a consistent hypercharge assignment of the form
(\ref{simpleform}) exists.  If so, we can then quickly determine the
corresponding value of $k_Y$ which describes the particular embedding,
and thereby determine the minimum value of $k_Y$ obtained.

We begin with the simplest case $\Delta=2$, and defer a discussion
of the higher-twist cases with $\Delta>2$ to the next section.
Given our masslessness, chirality, and moding restrictions, we
then find that for $\Delta=2$, the following are the only allowed
combinations for $(q_C,q_L)$:
\beqn
    (\rep{3},\rep{2}):&~~(q_C,q_L)=&
            (1/3,0),\,(-1/6,0),\,(-1/6,\pm 1/2)\nonumber\\
    (\rep{\overline{3}},\rep{1}):&~~(q_C,q_L)=&
            (-1/3,0),\,(-1/3,\pm 1/2),\, (1/6,0),\,
             (1/6,\pm 1/2)\nonumber\\
    (\rep{1},\rep{2})~~{\rm and}~~
    (\rep{1},\rep{1}):&~~(q_C,q_L)=&
             (0,\pm 1/2),\,(\pm 1/2,0),\,\pm(1/2,\pm 1/2)~.
\label{Deltatwocharges}
\eeqn
Thus, if we restrict our attention to the non-Higgs representations,
we see that we have in principle $4\cdot 6\cdot 6\cdot 8\cdot 8=9216$
different possible combinations of $(q_C,q_L)$ charges for the five
MSSM representations $\lbrace Q,u,d,L,e\rbrace$.
It is straightforward to automate an examination of each of these
possibilities,
and indeed one finds
that only for 16 combinations can
a consistent hypercharge $Y$ of the
form (\ref{simpleform}) be defined.
In fact, many of these 16 are related to each other
through trivial changes of sign, so that there are only
three different values of hypercharge normalization
$k_Y$ obtained.  These are
\beq
            k_Y ~=~  5/3,\, 11/3, ~{\rm and}~14/3~,
\eeq
with linear-combination coefficients $(A_1,A_2)=(-1/3,\pm 1/2)$,
$(2/3,\pm 1/2)$, and $(-1/3,\pm 1)$ respectively.
Thus we conclude that for $\Delta=2$ case and with $Y$ of the
form (\ref{simpleform}), we have
\beq
           k_Y ~\geq ~ 5/3~.
\eeq

Because of their importance, we now list the minimal
embeddings with $k_Y=5/3$.  It turns out that, modulo sign flips
and permutations, there are essentially two such inequivalent
embeddings with $k_Y=5/3$.  The first is the standard
``$SO(10)$ embedding'', so called due to the fact that the
charge vectors of the MSSM matter representations fill out the
weights of the $\rep{16}$ representation of $SO(10)$.\footnote{
    This includes the right-handed neutrino $N$ representation,
    which we have not listed in (\ref{sotenembedding}).}
In this embedding, the $(q_C,q_L)$ charges and a typical $\bQ$
vector for each of the MSSM representations are as follows:
\beqn
      Q:~~&(q_C,q_L)=(-1/6,0)&~~~~~\bQ=(-1/2,-1/2,+1/2,-1/2,+1/2)\nonumber\\
      u:~~&(q_C,q_L)=(+1/6,-1/2)&~~~~~\bQ=(1/2,1/2,-1/2,-1/2,-1/2)\nonumber\\
      d:~~&(q_C,q_L)=(+1/6,+1/2)&~~~~~\bQ=(1/2,1/2,-1/2,+1/2,+1/2)\nonumber\\
      L:~~&(q_C,q_L)=(+1/2,0)&~~~~~\bQ=(+1/2,+1/2,+1/2,-1/2,+1/2)\nonumber\\
      e:~~&(q_C,q_L)=(-1/2,+1/2)&~~~~~\bQ=(-1/2,-1/2,-1/2,+1/2,+1/2)~.~~~~
\label{sotenembedding}
\eeqn
This embedding corresponds to the solution $(A_1,A_2)=(-1/3,+1/2)$, which
successfully reproduces the required hypercharge assignments
(\ref{hypercharges}).

The second embedding with $k_Y=5/3$ bears no relation to the
$SO(10)$ embedding, yet shares the same value of $k_Y$.
In this embedding, the $(q_C,q_L)$ charges and a typical $\bQ$
vector for each of the MSSM representations are as follows:
\beqn
      Q:~~&(q_C,q_L)=(-1/6,0)&~~~~~\bQ=(-1/2,-1/2,+1/2,-1/2,1/2)\nonumber\\
      u:~~&(q_C,q_L)=(+1/6,-1/2)&~~~~~\bQ=(1/2,1/2,-1/2,-1/2,-1/2)\nonumber\\
      d:~~&(q_C,q_L)=(-1/3,0)&~~~~~\bQ=(0,0,-1,0,0)\nonumber\\
      L:~~&(q_C,q_L)=(0,-1/2)&~~~~~\bQ=(0,0,0,-1,0)\nonumber\\
      e:~~&(q_C,q_L)=(-1/2,+1/2)&~~~~~\bQ=(-1/2,-1/2,-1/2,1/2,1/2)~.~~~~
\label{minembedding}
\eeqn
This embedding also corresponds to the solution $(A_1,A_2)=(-1/3,+1/2)$, which
successfully reproduces the required hypercharge assignments
(\ref{hypercharges})
with these charge assignments.
It is a trivial matter to verify that in both of these embeddings with
$k_Y=5/3$,
the Higgs representations can also be successfully accommodated
 [{\it e.g.}, with charges $(q_C,q_L)_{H^\pm}=(0,\pm 1/2)$].

It is an interesting observation that two independent embeddings both
have the ``minimal value'' $k_Y=5/3$, and that all other possible
embeddings with $\Delta=2$ have only larger values of $k_Y$.
In this regard,
we emphasize that our chirality constraint has played a large role in
achieving this result.
For example, there {\it a priori}\/ exists an alternative embedding
which corresponds to an even smaller value of $k_Y$:
\beqn
      Q:~~&(q_C,q_L)=(+1/3,0)&~~~~~\bQ=(0,0,1,-1/2,1/2)\nonumber\\
      u:~~&(q_C,q_L)=(-1/3,+1/2)&~~~~~\bQ=(0,0,-1,1/2,1/2)\nonumber\\
      d:~~&(q_C,q_L)=(-1/3,-1/2)&~~~~~\bQ=(0,0,-1,-1/2,-1/2)\nonumber\\
      L:~~&(q_C,q_L)=(0,+1/2)&~~~~~\bQ=(0,0,0,0,1)\nonumber\\
      e:~~&(q_C,q_L)=(0,-1)&~~~~~\bQ=(0,0,0,-1,-1)~.
\label{nonchiralembedding}
\eeqn
For this set of $(q_C,q_L)$
charges, a
consistent hypercharge $Y$ can be found
corresponding to the solution $(A_1,A_2)=(1/6,-1/2)$.  This would
lead to a hypercharge normalization $k_Y=7/6\approx 1.17$, which is
substantially
smaller than $5/3\approx 1.67$.
However, we see from (\ref{nonchiralembedding}) that in this embedding,
the positron has the same charge vector as a gauge boson state
(in particular, the gauge boson perpendicular to $SU(2)_L$).
Hence this representation cannot be chiral:  either it is projected out
of the string spectrum altogether, or it appears with its complex conjugate.
It is for this reason that
the charge combination $(q_C,q_L)=(0,\pm 1)$ is
not listed as a possibility for the $(\rep{1},\rep{1})$
singlet representation in (\ref{Deltatwocharges}).  The only other combination
that was eliminated for the same reason is  $(q_C,q_L)=(1/3,\pm 1/2)$
for the $(\rep{3},\rep{2})$ representation, although its inclusion
does not lead to values $k_Y<5/3$ in any case.

Thus, we see that
the minimal embedding of the MSSM
gauge group and MSSM spectrum with the simplest moding $\Delta=2$
leads naturally to the minimum hypercharge normalization $k_Y=5/3$.
This is, of course, the value that is obtained in field-theoretic
GUT scenarios such as those making use of an $SO(10)$ embedding.
The important point here, however, is that we have made absolutely
no assumptions concerning any GUT scenario or embedding.  Rather,
we have employed only {\it string-based}\/ arguments, and found that
they too naturally lead to the same result.

As a final comment, we
note that our restriction to charge vector components $Q_i\in\IZ/2$
does {\it not}\/ mean that the corresponding model cannot involve
bosons with higher twists, or equivalently fermions with
multiperiodic boundary conditions.
This restriction merely implies that the MSSM chiral matter should not arise
from the sectors with higher twists.  In other words,
higher-twist sectors may appear in the string model,
but they should only have the effect of introducing extra non-MSSM matter
and/or implementing higher-twist GSO projections on the MSSM
sectors.
This is the certainly the case with most of the realistic
free-fermionic string models that have been examined in
the literature, such as those of Refs.~\cite{stringfsu5,slm,ALR}.
Even though these models contain higher twists, they do not
have chiral matter coming from those sectors.
Thus they are subject to our result as well, and since they
exhibit hypercharges of the ``minimal'' form (\ref{simpleform}),
they always have $k_Y\geq 5/3$.

\setcounter{footnote}{0}
\section{Hypercharge Embeddings:  Higher Twists and Higher \KM\ Levels}

In this section we apply the methods developed in the previous section
in order to consider two different extensions of those results.
The first extension involves higher twists, with charge modings $\Delta>2$,
and we will see in this case (unlike the case for $\Delta=2$),
there {\it do}\/ exist
isolated hypercharge embeddings with $k_Y<5/3$.  The second extension
involves the generalization to higher-level non-abelian gauge symmetries ---
in particular, to realizations of the non-abelian MSSM gauge group
$SU(3)\times SU(2)$ at levels $k_2,k_3>1$.  We shall find that this too
can have strong effects on the possible hypercharge embeddings.

\subsection{Higher Twists}

We first consider the cases with higher values of the moding $\Delta$.
As we have seen, restricting ourselves to
$\Delta=2$ leads naturally to the conclusion that $k_Y\geq 5/3$.
However, it is possible {\it a priori}\/
that the MSSM representations can arise from sectors
with higher twists, {\it e.g.}, sectors with $\Delta=4$ modings.
Thus such cases must be analyzed as well.
As we shall see, in these cases the number of possible
embeddings is much larger, and several isolated embeddings
within this class do have $k_Y<5/3$.  Whether these embeddings
can be realized within consistent string {\it models}\/ remains,
however, an open question.

At first glance, one might suspect that
chiral matter cannot possibly arise from sectors with
$\Delta>2$ modings;  the (ultimately faulty) line
of reasoning
would be that the presence of such higher-twist sectors
in a bosonic or fermionic construction corresponds to
Wilson lines in an orbifold formulation, and Wilson lines
can only introduce non-chiral matter.  If this were true,
our results from the last section would then be completely
general, and we would have shown that $k_Y\geq 5/3$ in all
level-one free-fermionic (or more generally, in all level-one
free-field) string models employing minimal hypercharge embeddings
of the form (\ref{simpleform}).  Unfortunately, however,
the correspondence between such
higher-twist sectors and Wilson lines is not as precise
as would be required for this type of claim,
and furthermore the assertion that Wilson lines do not
give chiral matter relies upon the compactification
manifold being sufficiently smooth.\footnote{
   We thank I. Antoniadis and E. Witten for discussions
   on this point.}
In particular, if there exist singularities
(such as exist in orbifold compactifications, for example),
then Wilson lines may give additional chiral matter.
Consequently, it is not possible to rule out the appearance
of chiral matter originating from higher-twist sectors.
Indeed, a simple example of a model in which
chiral matter {\it does}\/ arise in such higher-twist sectors
(in this case, sectors with $\Delta=6$) can be found in Ref.~\cite{Bailin}.
It is therefore also necessary to examine the cases with higher values of
$\Delta$, and in this subsection we shall analyze
the next-simplest case with $\Delta=4$.

Analyzing the set of possible hypercharge embeddings for $\Delta=4$
can be done in the same manner as for $\Delta=2$.
Imposing the masslessness, chirality,
and moding restrictions, it turns out that there are
16 possible combinations of $(q_C,q_L)$ charges
for the $(\rep{3},\rep{2})$ representation of $SU(3)\times SU(2)$,
25 possible combinations
for the $(\rep{\overline{3}},1)$ representation,
30 combinations for the $(\rep{1},\rep{2})$ representation,
and 36 for the $(\rep{1},\rep{1})$.
Including the Higgs $H^+$ and $H^-$ representations,
this yields potentially
$16\cdot (25)^2\cdot (30)^3 \cdot 36= 9.72\times 10^9$
different sets of charge vectors.
Of this number, however, only a relative few
are consistent with a hypercharge assignment
of the general form (\ref{simpleform}), and indeed
the overwhelming majority of those have corresponding
values of $k_Y$ greater than or equal to $5/3$.

Remarkably, however,
there are six isolated embeddings which
yield consistent hypercharge assignments with
$k_Y < 5/3$.
In order of increasing $k_Y$, the values obtained
are
\beq
  k_Y ~=~ { 53\over 48 },\, { 100 \over 81 },\,
     { 212 \over 147 },\, { 116\over 75 },\,
    { 77 \over 48 },\, { 236 \over 147 }~,
\eeq
which stretch within the
range $1.10 \leq k_Y \leq 1.61$.

\bigskip
\noindent {\it Cubic-Level Mass Terms}
\bigskip

In order to further restrict this set, it is possible to
impose one additional constraint.
As we have discussed in the previous section,
the MSSM hypercharge assignments are chosen
in part so that the mass terms
listed in (\ref{masstermconstraints}) will
be invariant under $U(1)_Y$.  This implies,
of course, that the sum of the hypercharges
of the three representations within each mass
term must be zero.  Within a {\it string-theoretic}\/
framework, however, such cubic mass terms (or
more generally, such direct couplings between
any number of fields)
are allowed
in the low-energy effective superpotential
only if the sum of the entire corresponding
charge {\it vectors}\/ vanishes.
Thus, not only must the hypercharge linear
combination $\sum_{\rm reps} \sum_i a_i Q_i$
vanish, but in fact the entire vector
$\sum_{\rm reps} \bQ$ must vanish.
This implies that the sums of the $q_C$ and $q_L$
quantum numbers must vanish separately.

While the cancellation of the hypercharge has
been assured by our hypercharge assignments
from the beginning, the cancellation of
the entire charge vectors is a completely
separate matter.
For example, it is possible to realize
an {\it effective}\/ mass term between three fields
$\Phi_i$, $\Phi_j$, and $\Phi_k$ even if
the sum of their charge vectors $\bQ_i+\bQ_j+\bQ_k$ does not cancel:
all that is required is a fourth field $\Phi_\ell$ whose
charge vector exactly cancels the sum, so
that the quartic coupling $\Phi_i\Phi_j\Phi_k\Phi_\ell$
is allowed.  Giving a VEV to $\Phi_\ell$ then introduces
the effective mass term
\beq
          \langle \Phi_\ell\rangle ~\Phi_i\Phi_j\Phi_k
\eeq
whose strength is set by the value of the VEV.
Such a mass term is said to arise at the quartic level in
the superpotential, and higher-order terms are also possible.
Thus, strictly speaking, it is not required for
the existence of an {\it effective}\/ mass term that the sum of the charge
vectors cancel;  indeed, this requirement holds only for mass terms
that arise at the cubic level in the superpotential.

There are, however, both stringy and phenomenological
reasons for distinguishing
between the cubic mass terms and the higher-order mass terms.
On the string side, if we do not require that the sum of the charge
vectors cancel, then we must instead demand that there simultaneously
exist in the spectrum an additional field (or product of fields) whose total
charge vector cancels $\bQ_i+\bQ_j+\bQ_k$, and which is also invariant under
the entire gauge symmetry $SU(3)\times SU(2)\times U(1)_Y$ (so that its
acquisition of a VEV does not break the MSSM gauge symmetry).
This is a difficult constraint to implement in a model-independent fashion,
and thus it is natural to demand (for simplicity) that {\it all}\/ mass terms
arise at only the cubic level of the superpotential.
Likewise, on the phenomenological side, it turns out that the order of such a
mass term
typically dictates the {\it scale}\/ of the corresponding coupling.  While the
scale of the cubic mass terms is typically fixed by $SL(2,\IC)$ invariance,
the higher-order mass terms  generally appear with inverse factors of the
Planck mass, and therefore yield effective mass terms that are increasingly
suppressed.
It is therefore reasonable to impose that at least one mass term be invariant
at
the cubic level.  Such a mass term will then be identified with the top quark
mass term, and constrain the allowed embeddings.  In a similar way, we might
additionally demand, for example, that {\it all}\/ of the heavy fermion mass
terms arise
at the cubic level, and so forth.
Thus, for both stringy and phenomenological reasons, it is natural
to impose the additional constraint on our embeddings
that they are consistent with each of the mass terms
in (\ref{masstermconstraints})
arising at the cubic level of the superpotential.

If we impose this additional constraint, we then find that
only two of our six embeddings with $k_Y<5/3$
survive.
These are the embeddings with
$k_Y=77/48\approx 1.604$
and $k_Y=236/147\approx 1.605$, and are as follows.
The $k_Y=77/48$ embedding consists of the charges
\beqn
      Q:~~&(q_C,q_L)=(-1/6,1/2)&~~~~~\bQ=(-1/2,-1/2,1/2,0,1)\nonumber\\
      u:~~&(q_C,q_L)=(-1/12,-3/4)&~~~~~\bQ=(1/4,1/4,-3/4,-3/4,-3/4)\nonumber\\
      d:~~&(q_C,q_L)=(5/12,-1/4)&~~~~~\bQ=(3/4,3/4,-1/4,-1/4,-1/4)\nonumber\\
      L:~~&(q_C,q_L)=(-1/4,-1/4)&~~~~~\bQ=(-1/4,-1/4,-1/4,-3/4,1/4)\nonumber\\
      e:~~&(q_C,q_L)=(1/2,1/2)&~~~~~\bQ=(1/2,1/2,1/2,1/2,1/2)\nonumber\\
    H^+:~~&(q_C,q_L)=(1/4,1/4)&~~~~~\bQ=(1/4,1/4,1/4,3/4,-1/4)\nonumber\\
    H^-:~~&(q_C,q_L)=(-1/4,-1/4)&~~~~~\bQ=(-1/4,-1/4,-1/4,-3/4,1/4)~,~~~
\label{lowembedding}
\eeqn
and corresponds to the hypercharge solution $(A_1,A_2)=(5/12,3/8)$.
Likewise, the $k_Y=236/147$ embedding consists of the
charges
\beqn
      Q:~~&(q_C,q_L)=(1/3,-1/4)&~~~~~\bQ=(0,0,1,-1/4,3/4)\nonumber\\
      u:~~&(q_C,q_L)=(-7/12,0)&~~~~~\bQ=(-1/4,-1/4,-5/4,0,0)\nonumber\\
      d:~~&(q_C,q_L)=(-1/12,-1/2)&~~~~~\bQ=(1/4,1/4,-3/4,-1/2,-1/2)\nonumber\\
      L:~~&(q_C,q_L)=(-1/4,1/4)&~~~~~\bQ=(-1/4,-1/4,-1/4,-1/4,3/4)\nonumber\\
      e:~~&(q_C,q_L)=(1/2,-1/2)&~~~~~\bQ=(1/2,1/2,1/2,-1/2,-1/2)\nonumber\\
    H^+:~~&(q_C,q_L)=(1/4,-1/4)&~~~~~\bQ=(1/4,1/4,1/4,1/4,-3/4)\nonumber\\
    H^-:~~&(q_C,q_L)=(-1/4,1/4)&~~~~~\bQ=(-1/4,-1/4,-1/4,-1/4,3/4)~,~~~
\label{otherembedding}
\eeqn
and corresponds to the hypercharge solution $(A_1,A_2)=(8/21,-3/7)$.
We find it remarkable that using only quarter-integer
components, there exist embeddings which
have the required forms
given in (\ref{fundreps}) and (\ref{singlets}),
which are simultaneously consistent with all of our
constraints ({\it e.g.}, masslessness,
chirality, and cubic-level mass terms),
and which also yield the correct
MSSM hypercharge assignments via linear combinations
corresponding to normalizations $k_Y<5/3$.

\bigskip
\noindent  {\it Existence of $k_Y<5/3$ String Models?}
\bigskip

Given the unique nature of these embeddings with $k_Y<5/3$,
the next step
is to attempt to realize them in a consistent string model.
After all, the embeddings that we have constructed
are meaningful only if they can be realized as the results
of actual string models.
This means that we must seek to construct a string model,
complete with corresponding GSO projections, in such a way that
all of the following conditions are met.  The first class of
conditions, of course,
are the model-independent constraints
which guarantee that our model is self-consistent:
these include
 worldsheet conformal anomaly cancellation,
 modular invariance,
 proper worldsheet supercurrent,
 proper spin-statistics, {\it etc}.
Next,
our second class of constraints are those that pertain
to the particular models we seek to construct.  In particular,
we must choose our GSO constraints in such a way that:
\begin{itemize}
\item  the model has $N=1$ spacetime supersymmetry;
\item  the appropriate gauge group $SU(3)\times SU(2)\times U(1)_Y$
        is realized;
\item  each of the desired MSSM representations in (\ref{representations})
         appears ({\it i.e.}, survives the GSO constraints);
\item  each such representation appears {\it chirally}\/
          ({\it i.e.}, the chiral conjugates of
          each MSSM representation must {\it not}\/ appear, and
           must therefore be eliminated by the GSO constraints); and
\item  the MSSM representations that survive the GSO constraints
         each have the same helicity.
\end{itemize}
Third, there are various additional constraints that
can be verified only after a particular model satisfying
the above constraints is constructed.  For example, we
must impose $U(1)_Y$ anomaly cancellation.
Although we are guaranteed by our original hypercharge assignments
that the MSSM representations by themselves are anomaly-free, one
must nevertheless verify that all of the {\it additional}\/ states
that necessarily appear in the full string spectrum are also
anomaly-free with respect to $U(1)_Y$.
Given the absence of the standard $SO(10)$ embeddings or
gauge symmetry, such anomaly-cancellations are not at all guaranteed.
Finally, additional constraints
that we might choose to impose for ``realistic'' models might
be constraints on the numbers of generations, and so forth.
Potential constraints arising from the appearance (or avoidance)
of fractionally-charged states will be discussed in the next
section.

To date, we have not succeeded in building
a self-consistent string model
which realizes these ``minimal'' $k_Y< 5/3$ embeddings
and which simultaneously satisfies the first and second
groups of constraints.
The primary difficulty arises in
realizing a set of GSO projections which is self-consistent
({\it i.e.}, which does not violate the first group of
constraints),  but which yields the necessary chirality
properties for the MSSM representations as listed above.
Thus, the existence of consistent string models
realizing these minimal $k_Y<5/3$ embeddings remains an open question.

\subsection{Higher Ka\v{c}-Moody Levels}

We now consider the possible hypercharge embeddings that
can be constructed when the $SU(3)$ and $SU(2)$ gauge factors
are realized at Ka\v{c}-Moody levels $k_3,k_2>1$.
There are many reasons why it is important to examine such situations.
For example, although the level-one models are the easiest to construct
within certain string formulations, the higher-level models
are also within the general moduli space of self-consistent models,
and moreover these models have certain appealing phenomenological
features (such as gauge group rank-reduction, and potentially smaller
sets of massless states and moduli).  It is therefore of interest,
for the purposes of gauge coupling unification,
to examine what values of $k_Y$
are possible for such models.
Second, a more practical reason for examining the higher-level
cases is that our analysis up to this point is highly dependent
on the fact that we have been working at level $k_2=k_3=1$.
Indeed, this choice directly determined the minimal gauge-group
embedding, which in turn dictated the possible matter embeddings,
and likewise the possible hypercharge embeddings.  Higher levels
for the non-abelian group factors
should therefore profoundly change the spectrum of possible
values of $k_Y$.  Third, as we have seen in Sect.~3,
for the purposes of gauge coupling unification, it is not
really $k_Y$ which is of interest, but rather the quotient $k_Y/k_2$.
Hence, even if (for some reason) one cannot realize
$k_Y<5/3$ in a realistic level-one models,
it may still be possible to realize
$k_Y<10/3$ in a level-{\it two}\/ model.
Finally, yet another reason for examining the cases with higher levels
has to do with charge integrality.  As we shall see in the next
section, the values of $k_Y$ that one can obtain in string models
containing only integrally charged color-neutral states
are tightly constrained by the values of $k_2$ and $k_3$.
This deep relation between $k_Y$ and the levels of the non-abelian
factors therefore suggests that a constructive examination of
possible hypercharge embeddings for higher-level models is in order.

In this section, we shall examine the case with $k_2=k_3=2$.
As we have seen in Sect.~3
[particularly in Fig.~\ref{rvsk2} and Eq.~(\ref{preferredvalues})],
models with $k_2=k_3=2$ are still within the phenomenologically preferred
ranges for gauge coupling unification.
Furthermore, we shall find that the $k_2=k_3=2$ case
is also relatively simple to analyze in the manner outlined in
the previous section, and we shall discuss the situations
for charge modings $\Delta=2$ and $\Delta=4$.
Extensions to higher levels and/or modings can then be handled in
a similar fashion.

\bigskip
\noindent {\it Gauge Group Embeddings}
\bigskip

We begin by considering the possible embeddings of the gauge groups
$SU(3)$ and $SU(2)$ realized at levels $k_3=k_2=2$.  From (\ref{lengthlevel}),
we see that whereas we previously required roots of
(length)$^2=2$, we now require roots of (length)$^2=1$.
Given this constraint, we then find that at least {\it four}\/ lattice
components
are necessary for embedding the root system of $SU(3)_2$,
with the two simple roots of $SU(3)$ chosen to have the coordinates
\beq
         (1,0,0,0) ~~~~~{\rm and}~~~~~~~ (-1/2,1/2,-1/2,1/2)~.
\label{higherlevelroots}
\eeq
As required, these two roots have a relative angle of 120$^\circ$.
The fact that $SU(3)$ has rank two implies that there are now
 {\it two}\/ directions orthogonal to the $SU(3)$ hyperplane,
and in the coordinate system set by (\ref{higherlevelroots})
these two directions can be taken to be
\beq
        Q_A~\equiv~ (0,1,1,0)~~~~~~{\rm and}~~~~~~
        Q_B~\equiv~ (0,0,1,1)~.
\eeq

For $SU(2)$, on the other hand,
the minimal embedding is particularly simple
at level $k_2=2$, and requires only {\it one}\/ lattice
direction.  We thus assign the single simple root of $SU(2)$ to have
lattice coordinate $(1)$.  The fact that such a one-dimensional
embedding is possible implies that there is no additional ``orthogonal''
component which must be introduced into this minimal embedding.

We see, then, that just as in the level-one case, the minimal
embedding of the root system of $SU(3)_2\times SU(2)_2$
requires a {\it five}\/-dimensional charge lattice, and we
shall once again denote these lattice components as $Q_i$, $i=1,...,5$,
where now $i=1,...,4$ denote the $SU(3)_2$ factor (and its
two orthogonal directions),
and $i=5$ denotes the $SU(2)_2$ factor.
Despite this superficial similarity to the level-one case,
however, there are some strong distinctions.
The most important of these concerns the appearance of {\it rank-cutting}\/
and of extra non-gauge chiral algebras.
As we can see from (\ref{centralcharges}),
the central charges of these gauge group factors
are respectively $c_{SU(3)}=16/5$ and $c_{SU(2)}=3/2$ at level two.
This means that although we are assigning two lattice directions
to $SU(3)$ and one lattice direction to $SU(2)$, in reality the
full $SU(3)$ conformal field theory must involve some additional
$c=6/5$ worth of worldsheet degrees of freedom
which are disconnected from the two lattice directions
we are considering here;  likewise the full $SU(2)$
conformal field theory requires an additional $c=1/2$ worth of
worldsheet degrees of freedom which are not reflected in
the gauge charge lattice.
This means that unlike the level-one case, certain worldsheet
degrees of freedom do not contribute to the charge lattice
of the model.  Indeed, it is in this manner that
the rank-cutting occurs, whereby the dimension of the total
charge lattice of the model is reduced.
In such theories, therefore, we see that the worldsheet excitations that
produce, {\it e.g.}, the non-Cartan gauge bosons of higher-level
gauge symmetries have two components:  some of the excitations
are among those worldsheet fields ({\it e.g.}, bosons or complex
fermions) which contribute to the charge lattice (thereby
creating a non-zero charge vector), and other excitations are
among those worldsheet fields (such as the {\it necessarily real}\/
fermions) that do not have a corresponding charge lattice.
It is for this reason that the charge vectors can be reduced in length
without altering the conformal dimensions of the corresponding
gauge boson states.
The same observations apply to the matter representations as well.

\bigskip
\noindent {\it Embeddings for Matter Representations}
\bigskip

Given the above embeddings for the root systems of $SU(3)_2$ and $SU(2)_2$,
it is then straightforward to deduce the corresponding embeddings for
the relevant matter representations.
We find
\beqn
       (\rep{3})_{(q_A,q_B)}~{\rm of}~SU(3): &~~~~ &
           \left\lbrace \matrix{ (0,1/3,-1/3,1/3) \cr
              (1/2,-1/6,1/6,-1/6)\cr
             (-1/2,-1/6,1/6,-1/6)\cr} \right\rbrace \nonumber\\
    &&~~~~~~~~~~~~~ ~~~~~~~+~ q_A\,(0,1,1,0) ~+~ q_B\,(0,0,1,1)\nonumber\\
       (\rep{1})_{(q_A,q_B)}~{\rm of}~SU(3): &~~~~ &
      q_A\,(0,1,1,0)~+~q_B\,(0,0,1,1)~\nonumber\\
    (\rep{2}) ~{\rm of}~ SU(2):&~~~~ &\pm (1/2)~\nonumber\\
    (\rep{1}) ~{\rm of}~ SU(2):&~~~~ & (0)~.
\label{fundrepshigherlevel}
\eeqn
As in the level-one case, we again have two quantum numbers $q_A$ and $q_B$
whose
values are {\it a priori}\/ arbitrary.
Unlike the level-one case, however, these two quantum numbers are now both
attached
to the $SU(3)$ representations, and the $SU(2)$ representations have no
remaining degrees of freedom.

\bigskip
\noindent {\it Conformal Dimensions}
\bigskip

We now consider the conformal dimensions of these representations (in order
to eventually enforce our masslessness conditions).
Unlike the level-one case, however,
the conformal dimension is no longer given by (\ref{chargedotproduct});
this occurs because in models with rank-cutting,
the conformal dimensions receive contributions not only from the
charge lattice excitations [which are tallied in (\ref{chargedotproduct})],
but also from the extra worldsheet degrees of freedom which
do not contribute to the charge lattice [and which are therefore
not reflected in (\ref{chargedotproduct})].
It is necessary, therefore, to use the more general expressions
(\ref{confdim}) when computing the conformal dimensions of each
given representation.
In this manner we then determine the conformal dimensions
for the relevant representations of $SU(3)_2\times SU(2)_2$:
\beqn
    (\rep{3},\rep{2})_{(q_A,q_B)}~:&&~~~~~h~=~109/240 ~+ ~{q_A}^2 ~+~
       {q_B}^2\nonumber\\
    (\rep{\overline{3}},\rep{1})_{(q_A,q_B)}~:&&~~~~~
                      h~=~4/15 ~+ ~{q_A}^2 ~+~ {q_B}^2\nonumber\\
    (\rep{1},\rep{2})_{(q_A,q_B)}~:&&~~~~~h~=~3/16 ~+ ~{q_A}^2 ~+~
       {q_B}^2\nonumber\\
    (\rep{1},\rep{1})_{(q_A,q_B)}~:&&~~~~~h~=~{q_A}^2 ~+~ {q_B}^2~.
\label{confdimshigherlevel}
\eeqn
Of course, the allowed values of $(q_A,q_B)$ for each representation
are then constrained by the requirement that the corresponding
conformal dimension must be less than one, and likewise by
whatever moding requirements we wish to impose.
We also impose the chirality condition discussed in the previous section.
In any case, however, the ultimate constraint on values of $(q_A,q_B)$
for each representation
comes from the requirement that a consistent
hypercharge assignment for each of the MSSM representations
must be simultaneously realizable.

\bigskip
\noindent {\it Hypercharge Embeddings}
\bigskip

We now approach the question of the corresponding hypercharge embeddings.
As before, we consider only the ``minimal'' embedding of the form
(\ref{simpleform}), and determine the solutions for which the corresponding
values of $k_Y$ are in the appropriate range.

Given our hypercharge embedding of the form (\ref{simpleform}), we
immediately observe that for the higher-level matter embeddings
in (\ref{fundrepshigherlevel}),
we now must have
\beq
          a_1~=~a_5~=~0 ~~~~~~~{\rm and}~~~~~~~ a_2~+~a_4~=~a_3~.
\eeq
These constraints are necessary in order to ensure that
the same hypercharge value $Y$ is obtained for each charge vector
within a single representation.
Thus, we see that once again there are only two independent
coefficients, $a_2$ and $a_4$, which are unconstrained.
As expected, this renders the hypercharge
assignment independent of the particular $SU(3)$ or $SU(2)$
representations $\rep{R}$ or $\rep{R'}$,
so that $Y$ depends on only the $(q_A,q_B)$
quantum numbers:
\beq
        Y \, (\rep{R}_{(q_A,q_B)}, \rep{R'})
     ~=~ a_2\,q_A ~+~ (a_2+a_4)\,(q_A+q_B) ~+~ a_4\,q_B~.
\label{Yatwoafour}
\eeq
For any such solution $(a_2,a_4)$, the
corresponding value of $k_Y$ is then given by
\beq
      k_Y~=~ 2\,\sum {a_i}^2 ~=~
            2\,\biggl\lbrack {a_2}^2 + (a_2+a_4)^2 + {a_4}^2\biggr\rbrack~.
\eeq
The task therefore remains to choose a particular value of moding $\Delta$,
and to determine the possible corresponding embeddings for which
a hypercharge may be consistently defined.

We first present results for the moding $\Delta=2$.
Choosing this moding amounts to the restriction that
$q_A$ and $q_B$ be in the set $1/6 + \IZ/2$ for the triplet $\rep{3}$
representation of $SU(3)$,
or in the set $\IZ/2$ for the singlet representation of $SU(3)$.
Imposing the masslessness and chirality constraints then yields
a finite set of possibilities of $(q_A,q_B)$ for each representation.
This ultimately yields several hundred thousand combined possibilities
for the charges $(q_A,q_B)$ of the representations $(Q,u,d,L,e,H^+)$.
However, we find that only for a relatively small set are consistent
hypercharge assignments possible, and indeed the values
of $k_Y$ to which they correspond are:
\beq
      k_Y ~=~ {4\over 3},~ {13\over 3},~ {28\over 3},~ {52\over 3}~.
\eeq
The embedding with $k_Y=52/3$
is in fact unique ({\it i.e.}, there is only one embedding with this
value), and it does not satisfy the cubic-level mass-term constraints
discussed in the previous section.
By contrast,
among the embeddings corresponding to each other value of $k_Y$,
there are some which satisfy these cubic-level mass-term
constraints, and some which do not.

It is clear from these results, then, that the spectrum of $k_Y$ values
realizable in such embeddings is quite limited;  moreover, the value $k_Y=4/3$
is special in that is already less than 5/3 (with the value of $k_Y/k_2$
less than one, no less!).
It turns out that there are two independent hypercharge embeddings
which take this value of $k_Y$ and which also satisfy all of the cubic-level
mass-term constraints.  These embeddings are respectively
\beqn
      Q:~~&(q_A,q_B)=(-1/3,1/6)&~~~~~\bQ=(0,0,-1/2,1/2,1/2)\nonumber\\
      u:~~&(q_A,q_B)=(1/3,1/3)&~~~~~\bQ=(0,0,1,0,0)\nonumber\\
      d:~~&(q_A,q_B)=(1/3,-2/3)&~~~~~\bQ=(0,0,0,-1,0)\nonumber\\
      L:~~&(q_A,q_B)=(1/2,0)&~~~~~\bQ=(0,1/2,1/2,0,1/2)\nonumber\\
      e:~~&(q_A,q_B)=(-1/2,-1/2)&~~~~~\bQ=(0,-1/2,-1,-1/2,0)\nonumber\\
    H^+:~~&(q_A,q_B)=(0,-1/2)&~~~~~\bQ=(0,0,-1/2,-1/2,1/2)\nonumber\\
    H^-:~~&(q_A,q_B)=(0,1/2)&~~~~~\bQ=(0,0,1/2,1/2,-1/2)~,
\label{embeddinghigherlevelone}
\eeqn
and
\beqn
      Q:~~&(q_A,q_B)=(-1/3,1/6)&~~~~~\bQ=(0,0,-1/2,1/2,1/2)\nonumber\\
      u:~~&(q_A,q_B)=(5/6,-1/6)&~~~~~\bQ=(0,1/2,1,-1/2,0)\nonumber\\
      d:~~&(q_A,q_B)=(-1/6,-1/6)&~~~~~\bQ=(0,-1/2,0,-1/2,0)\nonumber\\
      L:~~&(q_A,q_B)=(0,1/2)&~~~~~\bQ=(0,0,1/2,1/2,1/2)\nonumber\\
      e:~~&(q_A,q_B)=(-1/2,-1/2)&~~~~~\bQ=(0,-1/2,-1,-1/2,0)\nonumber\\
    H^+:~~&(q_A,q_B)=(-1/2,0)&~~~~~\bQ=(0,-1/2,-1/2,0,1/2)\nonumber\\
    H^-:~~&(q_A,q_B)=(1/2,0)&~~~~~\bQ=(0,1/2,1/2,0,-1/2)~,
\label{embeddinghigherleveltwo}
\eeqn
both of which correspond to the solution $a_2=a_4= -1/3$.
The fact that there are no embeddings with values of $k_Y$
in the range $5/3\leq k_Y<10/3$
is unfortunate, however, as we {\it a priori}\/ seek values
$k_Y\approx 2.8$ for these $k_2=2$ embeddings.

This deficiency can be overcome, however, by considering
the more general moding $\Delta=4$, and indeed this
introduces additional potential hypercharge embeddings
which satisfy all of the cubic-level mass-term constraints
and which do have $k_Y/k_2<5/3$.
In fact, the particular set of values of $k_Y$ to which
such embeddings correspond is remarkably limited.
We find, for example, that for $\Delta=4$, there exist such potential
embeddings with only the following values of $k_Y<5/3$:
\beq
        k_Y~=~
        {4\over 3},~
        {112\over 81},~
        {208\over 147},~
        {112\over 75}~.
\eeq
Similarly, within the range $5/3\leq k_Y<10/3$, the only values
of $k_Y$ for which such embeddings exist are the following:
\beq
        k_Y~=~
        {16\over 9},~
        {7\over 3},~
        {208\over 75},~
        {28\over 9}.
\eeq

We see from these results, then, that
there exist embeddings with $k_Y\approx 2.8$ which
are potentially realizable in $k_2=2$ string models,
and which satisfy all of our constraints (including
all of the cubic-level mass-term constraints).
In particular, these are the embeddings
with $k_Y=208/75\approx 2.8$, so that
$k_Y/k_2\approx 1.4$.
It turns out that there are three distinct embeddings
with this value of $k_Y$ (modulo trivial sign flips
and permutations).
The first corresponds to the hypercharge solution
$(a_2,a_4)=(4/15,2/3)$, and is
\beqn
      Q:~~&(q_A,q_B)=(-1/12,1/6)&~~~~~\bQ=(0,1/4,-1/4,1/2,1/2)\nonumber\\
      u:~~&(q_A,q_B)=(1/3,-2/3)&~~~~~\bQ=(0,0,0,-1,0)\nonumber\\
      d:~~&(q_A,q_B)=(-1/6,1/3)&~~~~~\bQ=(0,-1/2,1/2,0,0)\nonumber\\
      L:~~&(q_A,q_B)=(-3/4,1/4)&~~~~~\bQ=(0,-3/4,-1/2,1/4,1/2)\nonumber\\
      e:~~&(q_A,q_B)=(1/2,1/4)&~~~~~\bQ=(0,1/2,3/4,1/4,0)\nonumber\\
    H^+:~~&(q_A,q_B)=(-1/4,1/2)&~~~~~\bQ=(0,-1/4,1/4,1/2,1/2)\nonumber\\
    H^-:~~&(q_A,q_B)=(1/4,-1/2)&~~~~~\bQ=(0,1/4,-1/4,-1/2,-1/2)~.
\label{goodembeddingone}
\eeqn
The remaining two $k_Y=208/75$ embeddings, by contrast,
correspond to the different hypercharge solution $(a_2,a_4)=(-14/15,2/3)$,
and are
\beqn
      Q:~~&(q_A,q_B)=(-1/12,1/6)&~~~~~\bQ=(0,1/4,-1/4,1/2,1/2)\nonumber\\
      u:~~&(q_A,q_B)=(1/3,-2/3)&~~~~~\bQ=(0,0,0,-1,0)\nonumber\\
      d:~~&(q_A,q_B)=(-1/6,1/3)&~~~~~\bQ=(0,-1/2,1/2,0,0)\nonumber\\
      L:~~&(q_A,q_B)=(1/2,1/4)&~~~~~\bQ=(0,1/2,3/4,1/4,1/2)\nonumber\\
      e:~~&(q_A,q_B)=(-3/4,1/4)&~~~~~\bQ=(0,-3/4,-1/2,1/4,0)\nonumber\\
    H^+:~~&(q_A,q_B)=(-1/4,1/2)&~~~~~\bQ=(0,-1/4,1/4,1/2,1/2)\nonumber\\
    H^-:~~&(q_A,q_B)=(1/4,-1/2)&~~~~~\bQ=(0,1/4,-1/4,-1/2,-1/2)~
\label{goodembeddingtwo}
\eeqn
and
\beqn
      Q:~~&(q_A,q_B)=(-1/12,1/6)&~~~~~\bQ=(0,1/4,-1/4,1/2,1/2)\nonumber\\
      u:~~&(q_A,q_B)=(7/12,1/12)&~~~~~\bQ=(0,1/4,1,-1/4,0)\nonumber\\
      d:~~&(q_A,q_B)=(-5/12,-5/12)&~~~~~\bQ=(0,-3/4,-1/2,-3/4,0)\nonumber\\
      L:~~&(q_A,q_B)=(1/4,-1/2)&~~~~~\bQ=(0,1/4,-1/4,-1/2,1/2)\nonumber\\
      e:~~&(q_A,q_B)=(-3/4,1/4)&~~~~~\bQ=(0,-3/4,-1/2,1/4,0)\nonumber\\
    H^+:~~&(q_A,q_B)=(-1/2,-1/4)&~~~~~\bQ=(0,-1/2,-3/4,-1/4,1/2)\nonumber\\
    H^-:~~&(q_A,q_B)=(1/2,1/4)&~~~~~\bQ=(0,1/2,3/4,1/4,-1/2)~.
\label{goodembeddingthree}
\eeqn

It is clear, then, that within the context of higher-level
realizations of the $SU(2)$ and $SU(3)$ gauge groups,
there exist potential hypercharge embeddings with $k_Y/k_2$
in the desired range $k_Y/k_2 \approx 1.4$.
Furthermore, the freedom to build embeddings beyond
those of the form (\ref{simpleform}) should also enable
other nearby values of $k_Y$ to be reached, although we
would {\it a priori}\/ expect the inclusion of additional
lattice components in the hypercharge linear combination
to tend to increase the corresponding values of $k_Y$.
Nevertheless, the spectrum of allowed values of $k_Y$ within just the
``minimal'' embeddings of the form (\ref{simpleform})
is sufficiently diverse
to suggest that within the class of higher-twisted or higher-level
string models,
various phenomenologically interesting values of $k_Y$
can be obtained without difficulty.
Of course, it remains an open question as to whether
these special embeddings can be realized within
actual self-consistent string models.
The existence of these special embeddings
should nevertheless provide a useful starting point in
this endeavor.


\setcounter{footnote}{0}
\section{Charge Quantization Constraints}

In the previous two sections, we analyzed the constraints
on the values of $k_Y$ that arise for various ``minimal embeddings''
of the MSSM gauge group.
In this way we were able to obtain certain interesting
results.  For example,
by proving the non-existence of $k_Y < 5/3$ embeddings
for $\Delta=2$ charge modings and level-one
$SU(2)$ and $SU(3)$ gauge factors,
we were able to prove that no such models utilizing
such minimal embeddings can have $k_Y<5/3$.
Likewise, in the cases of more general
modings and higher \KM\ levels,
we were able to construct potentially viable embeddings
that do have $k_Y<5/3$ (or more generally $k_Y/k_2<5/3$).
These embeddings should therefore prove useful in the construction
of phenomenologically interesting $k_Y<5/3$ string models.

As we discussed in Sect.~4.1, however, such an approach has the
disadvantage that it is beyond the reach of certain powerful
string consistency constraints such as modular invariance.
In this section, therefore, we shall now follow a somewhat orthogonal
approach to constraining the MSSM levels $(k_Y,k_2,k_3)$, one
which incorporates modular invariance at an early stage.
As we shall find, this method will allow us to correlate the
combinations of \KM\ levels with the appearance of
fractionally charged states in the string spectrum, and thereby
constrain the phenomenologically preferred combinations of levels
$(k_Y,k_2,k_3)$ for which such states do not appear.

\subsection{The Method}

As originally shown by Schellekens \cite{Schellekens},
an interesting set of restrictions on the \KM\ levels $(k_Y,k_2,k_3)$ arises
by imposing charge-quantization conditions on the asymptotic states
of the theory in the infrared.
These conditions are motivated by the observation that in the observed MSSM
particle spectrum, there is a strong correlation between the
allowed combinations of $SU(3)\times SU(2)\times U(1)$ representations:
all color singlets have integer electric
charge (in units of the electron charge).
Indeed, there are also quite stringent experimental constraints on the
existence of massive fractionally charged states, stemming from the fact
that despite various experimental efforts, fractionally charged states
have not been seen~\cite{fractionalexpt}.
In this section, therefore, we will consider the consequences of imposing
a charge integrality condition on the allowed states in string theory.
First, as we shall see, if we simply demand that
that the spectrum of a given string model does not contain
fractionally charged states {\it at all}\/,
then it is straightforward to show that we must have in fact have
$k_Y\geq 5/3$ for level-one models ({\it i.e.}, models with $k_2=k_3=1$).
This result is essentially contained within Schellekens' original analysis.
However, string models generically {\it will}\/ contain fractionally charged
states;  such states are then presumed to be {\it bound}\/ into color
singlets under the influence of some extra hidden confining gauge symmetries
(or ``hypercolor'' groups) beyond those of the MSSM.
We shall therefore generalize the Schellekens analysis for the cases
of arbitrary hypercolor groups beyond that of color $SU(3)$, and require
only the weaker constraint that all of the string states which appear
can be bound into color singlets under the hypercolor interactions.
This will then enable us
to determine which more general combinations of \KM\ levels (and in
particular which more general values of $k_Y$) are allowed.
In a similar analysis, we shall also consider a slightly different
generalization
of the charge integrality constraints in order to determine the values of $k_Y$
that can arise when only a restricted set of fractional charges
are permitted to appear in the spectrum.
Note that such charge-integrality constraints
on the \KM\ levels have also been previously considered for the purposes
of analyzing certain specific string models \cite{ellis,anton}.

The starting point of our analysis is a theorem
proved by Schellekens~\cite{Schellekens}\ utilizing the properties of
simple (or ``bonus-symmetry'') currents that arise in many rational conformal
field theories
(RCFT's)~\cite{schellyank,keni}.  Recall that a rational conformal
field theory is one that contains only finitely many primary fields.
In such a theory, a simple current $J$ is then defined as a primary field
that has a one-term fusion rule with all primary fields $\phi_p$
in the theory, so that we obtain a fusion rule of the simple form
\beqn
        J ~\times~ \phi_p ~=~ \phi_{p'}~.
\label{uniquefusion}
\eeqn
Note that for each primary field $\phi_p$, the resulting field $\phi_{p'}$ is
unique.
Thus, when acting on primary fields $\phi_p$, each simple current $J$ generates
a
cyclic action of order $N$, where $N$ is the smallest integer such that $J^N =
       {\bf 1}$ under
fusion.
In the following, we shall denote the identity current as $J_0\equiv {\bf 1}$.
Note that the order $N$ must be finite due to our restriction to RCFT's.
Also note that, by construction,  the identity operator $J_0$ is always a
simple current
of order $N=1$.  With these definitions, we then have:

 {\it Theorem.}\cite{Schellekens}~  For any rational unitary conformal field
theory, consider the partition function
\beqn
     Z(\tau, \overline{\tau}) ~=~ \sum_{m,\tilde{n}} \,\chi_m(\tau)\,
      M_{m \tilde{n}}\, \tilde{\chi}_{\tilde{n}}(\overline{\tau})~,
\label{partition}
\eeqn
where $M_{m\tilde{n}}$ is a matrix of positive integers, and where
the left and right-moving sets of characters $\chi$ and $\tilde{\chi}$
can be different. Denote the primary fields corresponding to a left-right
combination of characters as $\phi_{m\tilde{n}}$. Suppose that $p$
and $q$ are labels of simple currents, such that $\phi_{p\tilde{q}}$
is local with respect to all other primary fields $\phi_{m\tilde{n}}$.
Let us also suppose that $M_{m\tilde{n}}\neq 0$, so that
the primary fields $\phi_{m\tilde{n}}$ are present in the theory.
Then the modular invariance of the partition
function $Z(\tau, \overline{\tau})$ {\it requires}\/ that
$M_{p\tilde{q}}\neq0$.
Hence the primary field $\phi_{p\tilde{q}}$ must also be present in the theory.

In this theorem, the condition of locality is the statement that in
operator products of $J(z)$ with any primary field $\phi(w)$, we have
\beqn
     J(z) \phi(w) ~\sim~ { \phi'(w) \over (z-w)^{\alpha}}
                  ~+~ ...
\label{opelocal}
\eeqn
where the exponent $\alpha \equiv h(J) + h(\phi) - h(\phi')$ is an integer.
(Likewise, we have similar relations for the anti-holomorphic OPE's;  these
will not be written in what follows.)
Note that the {\it monodromy}\/ of the current $J$ with respect to a given
field $\phi$ is defined as
\beqn
 {\rm mono}(J,\phi) ~=~ h(J) + h(\phi) - h(\phi')~~({\rm mod}~1)
\label{monodef}
\eeqn
where $\phi'$ is the primary field obtained in (\ref{opelocal}).
Thus, locality is equivalent to requiring that ${\rm mono}(J,\phi)\equiv 0$ for
all $\phi$.  Note also that
\beqn
 {\rm mono}(J\times J',\phi) ~=~ {\rm mono}(J,\phi) ~+~ {\rm mono}(J',\phi) ~.
\eeqn
We shall have occasion to use this result later.

An important point to notice is that the conditions stated
in the above theorem are not sufficient
to guarantee the existence of a non-trivial modular invariant partition
function
including $\phi_{p\tilde{q}}$.
Indeed, there is an extra condition that must
be satisfied by $\phi_{p\tilde{q}}$:
its holomorphic and anti-holomorphic conformal dimensions $(h(p),
\overline{h}(q))$
must differ by an integer in order for the partition function to be invariant
under the
modular transformation $T:\tau \rightarrow \tau + 1$.  As we shall see, this
consistency condition on the conformal dimension of the simple
currents will lead to a condition on the levels of the \KM\ algebras,
for the conformal dimensions of the simple currents depend on the levels $k_i$.
Thus, the logic of Schellekens argument is as follows.
First, we demonstrate that given the charge-quantization
condition (which is to be satisfied by all of the string states), it is
possible
to choose
a combination $\tilde{J}$ of simple currents
such that
the monodromy of $\tilde J$ with respect to these states
is {\it automatically}\/ integral.
By the theorem quoted above, this implies
that $\tilde{J}$ must appear as a primary field in the theory.
Second, we then demand that the holomorphic and anti-holomorphic
conformal dimensions of $\tilde J$
differ by integers.
This will then enable us to
obtain a relation among the allowed \KM\ levels $k_i$.

\subsection{Simple Currents for Classical Lie Algebras}

Towards this end,
our first concern is with the simple currents that occur in \KM\
algebras. For the \KM\ algebras associated with the classical
Lie algebras (which are the only cases we will consider), the enumeration of
the non-trivial simple currents at level $k$ is
straightforward~\cite{schellyank,keni,ver,fuchs}.
Of course, for each case, the identity $J_0\equiv {\bf 1}$ is always a simple
current.  We shall now list the remaining non-trivial currents that arise for
each
group.

 \underbar{$SU(n+1)_k:$}~~
For $A_n\equiv SU(n+1)$ at level $k$, there are $n+1$ simple currents
which we denote $J_A^{(m)}$ with $m=0,...,n$.
The case with $m=0$ is the identity current $J_0$, and
the remaining currents $J_A^{(m)}$
with $m=1,...,n$ are specified by the
$SU(n+1)$ representations with Dynkin indices
$(0,...,0,k,0,...,0)$ where the non-zero entry is
the $m^{\rm th}$ index.
Note that under
fusion, we have $(J_A^{(1)})^{n+1} = J_0$.
The conformal dimensions of these currents are
\beqn
     h(J_A^{(m)}) ~=~ {k\, m \,(n+1-m) \over 2(n+1)}~,
\label{suh}
\eeqn
in accordance with the general expression (\ref{confdim}).
The other information that we will need is the monodromy of these currents
with any primary field $\phi$ transforming in an arbitrary irreducible
representation $R$ of $SU(n+1)_k$.
This monodromy is given by
\beqn
       {\rm mono}(J_A^{(m)},R) ~=~ {m \,c_{su(n+1)}(R) \over n+1}~,
\label{sumono}
\eeqn
where $c_{su(n+1)}(R)$ is the so-called congruence class [or ``$(n+1)$-ality'']
of the
representation $R$ in $SU(n+1)$.  This congruence class is an
integer modulo $(n+1)$, and is defined
in terms of the Dynkin indices $a_{(j)}$ of $R$ as
\beqn
         c_{su(n+1)}(R) ~\equiv~ \sum_{j=1,...,n} j \,a_{(j)}~.
\label{sunconjc}
\eeqn

 \underbar{$SO(2n+1)_k:$}~~
For $B_n\equiv SO(2n+1)$ at level $k$, there is just one non-trivial
simple current $J_B$.  This current has Dynkin indices $(k,0,...,0)$,
and  its conformal dimension is
\beqn
          h(J_B) ~=~ k/ 2~.
\label{sooddh}
\eeqn
Likewise, its monodromy with respect to any representation
$R$ of $SO(2n+1)_k$ is
\beqn
        {\rm mono}(J_B,R) ~=~ {1\over 2}\, c_{so(2n+1)}(R) ~,
\label{sooddmono}
\eeqn
where the congruence class $c_{so(2n+1)}(R)$ is defined ${\rm mod}~2$,
and just labels whether $R$ is vector-like (with $c= 0$) or
spinor-like (with $c= 1$).

 \underbar{$Sp(2n)_k:$}~~
For $C_n\equiv Sp(2n)$ at level $k$, there is again just one non-trivial
simple current $J_C$;  this current has Dynkin indices $(0,...,0,k)$,
and has  conformal dimension
\beqn
         h(J_C) ~=~ {nk\over 4}~.
\label{sph}
\eeqn
Likewise, the monodromy of this current with respect
to any representation $R$ of $Sp(2n)_k$ is
\beqn
     {\rm mono}(J_C,R) ~=~ {1\over 2}\, c_{sp(2n)}(R) ~,
\label{spmono}
\eeqn
where the congruence class $c_{sp(2n)}(R)$ is again defined ${\rm mod}~2$,
and now corresponds to the reality or pseudoreality of the representation $R$
(with $c=0$ for real representations, and $c=1$ for pseudo-real
representations).

 \underbar{$SO(2n)_k:$}~~
Finally, the $D_n\equiv SO(2n)$ algebras at level $k$ possess
three non-trivial simple currents, $J_D^{(v)}=(k,0,...,0)$,
$J_D^{(s)}=(0,...,0,k)$,
and $J_D^{(c)} =(0,...,0,k,0)$, corresponding to the vector, spinor, and
conjugate-spinor
representations.
These currents have conformal dimensions
\beq
	     h(J_D^{(v)})~=~  {k\over 2}~,~~~~~
             h(J_D^{(s)})~=~  {kn\over 8}~, ~~~~~
             h(J_D^{(c)})~=~  {kn\over 8}~,
\label{soevendim}
\eeq
and satisfy the fusion rules $J_D^{(v)} \times J_D^{(s)} = J_D^{(c)}$
and cyclic permutations.
Unlike the previous cases, however, the structure of the
monodromy of these currents with any given representation
$R$ of $SO(2n)_k$ depends on whether $n$ is odd or even.  This distinction
ultimately arises because the center of $SO(2n)$ is $\IZ_4$ if $n$ is odd,
but $\IZ_2 \otimes \IZ_2$ if $n$ is even.
For $n$ odd, we have
\beqn
        {\rm mono}(J_v,R) &=&  {1\over 2}\, c_{so(4m+2)}(R) ~, \nonumber\\
        {\rm mono}(J_s,R) &=&  {1\over 4}\, c_{so(4m+2)}(R) ~,  \nonumber\\
        {\rm mono}(J_c,R) &=&  {\rm mono}(J_v,R)~+~ {\rm mono}(J_s,R)~,
\label{somonoo}
\eeqn
where the congruence class $c_{so(4m+2)}(R)$ is defined
$({\rm mod}~4)$ in terms of the Dynkin indices of $R$
by the explicit sum
\beqn
   c_{so(4m+2)}(R)=
   2a_{(1)} + ... + 2a_{(2m-3)} + (2m-1)a_{(2m)} + (2m+1)a_{(2m+1)}.
\label{oddcc}
\eeqn
Note that although the congruence class is defined mod 4,
all of the monodromies in (\ref{somonoo}) are nevertheless defined mod 1.
Likewise, for $n$ even, we have
\beqn
	{\rm mono}(J_v,R) &=&  {1\over 2}\, c^{(1)}_{so(4m)}(R) ~, \nonumber\\
        {\rm mono}(J_s,R) &=&  {1\over 2}\, c^{(2)}_{so(4m)}(R) ~, \nonumber\\
        {\rm mono}(J_c,R) &=&  {\rm mono}(J_v,R)~+~ {\rm mono}(J_s,R)~,
\label{somonoe}
\eeqn
where now $c^{(i)}_{so(4m)}(R)$ ($i=1,2$) are the two components, both
defined mod~$2$, of the charge conjugacy vector describing $\IZ_2 \otimes
\IZ_2$.
Explicitly, we have
\beqn
       c^{(1)}_{so(4m)}(R) &=& a_{(2m-1)} + a_{(2m)}\nonumber\\
       c^{(2)}_{so(4m)}(R) &=& a_{(1)} + a_{(3)} + ... + a_{(2m-3)} +
       (m-1)a_{(2m-1)} + ma_{(2m)}.
\label{evencc}
\eeqn

\subsection{Charge Integrality Constraints with Confining Group}

Given the above simple currents and monodromies, the first step in
our derivation of the integrality constraint is
to express the charge-quantization condition
in terms of conjugacy classes.
Let us generally take our low-energy gauge group
to be of the form
$G \times SU(3) \times SU(2) \times U(1)_Y$, where  $G$ is assumed to
be semi-simple.
Likewise, let us take
each of the states $\psi_i$
of our theory (whether massless or massive) to transform
in some representation $(R_i,r_i,\tilde{r}_i,Y_i)$ of this group.
The condition that all $SU(3)$-color and $G$-hypercolor
singlets be integrally charged with respect to the electromagnetic
charge operator $Q=T_3 + Y$ then takes the form
\beqn
     \alpha_G(R_i) + {c_{su(3)}(r_i)\over 3} + { c_{su(2)}(\tilde{r}_i)\over 2}
     + Y_i ~\equiv~ 0~~~~({\rm mod}~1)
\label{quantform}
\eeqn
for each and every state $\psi_i$ in the theory.
Here the expression $\alpha_G(R)$ is an additive group-dependent
quantity which is calculated as follows.
For each gauge-group factor of $SU(n+1)$, $SO(2n+1)$, $Sp(2n)$,
$SO(4m+2)$, or $SO(4m)$ which is present in $G$, $\alpha_G(R)$ is obtained
by adding together the corresponding factors:
\beqn
     SU(n+1):&~~~
     z\,c_{su(n+1)}(R)/(n+1) ~~~~     ~~~~&{\rm for}~z=0,...,n\nonumber\\
     SO(2n+1):&~~~
     z\,c_{so(2n+1)}(R)/2  ~~~~     ~~~~ &{\rm for}~z=0,1\nonumber\\
     Sp(2n):&~~~
     z\,c_{sp(2n)}(R)/2  ~~~~     ~~~~  &{\rm for}~z=0,1\nonumber\\
     SO(4m+2):&~~~
     z\,c_{so(4m+2)}(R)/4  ~~~~    ~~~~  &{\rm for}~z=0,1,2,3\nonumber\\
     SO(4m):&~~~
     \bigl(z_1 \,c^{(1)}_{so(4m)}(R)/2 ~+~ z_2 \,c^{(2)}_{so(4m)}(R)
        \bigr)/2 ~~~~~~&{\rm for}~z_1,z_2 \in \lbrace 0,1\rbrace ~.\nonumber\\
   && ~~~~
\label{alphaparts}
\eeqn
The presence of the various integers $z$, $z_1$, and $z_2$ in these formulas
expresses the fact that, for a given confining group, there is more than one
possible assignment of charges to the ``quarks'' that leads to only
integrally charged bound states.
In particular, if the chosen value of $z$ divides the order of
the center of the group (with common divisor $d$),
then the minimally charged bound state has charge $d$ in
units of the electron charge.  If, however, $z$ does not divide the
order, then the charges of the ``hypercolor quarks'' can be reassigned in
a physically distinct way such that the minimal charge of the bound
states remains one.
These arbitrary factors of $z$  are therefore inserted
into (\ref{alphaparts}) in order to reflect the freedom allowed
by such reassignments of the electric charges.

Note that the fact that the condition (\ref{quantform}) is necessary
follows simply from the fact that singlets always live in the trivial
conjugacy class.  Likewise, the fact that this condition is sufficient follows
from a
consideration of the invariants of the classical Lie algebras.
Also note that for the sake of notational brevity,
we have dropped the species label $i$
in (\ref{alphaparts}) and in what follows.

The next step is
to construct a combination $\tilde{J}$
of the individual \KM\ simple
currents in such a way that $\tilde{J}$ is automatically local with respect to
all states in the theory by virtue of the quantization
condition (\ref{quantform}). For simplicity, let us suppose for the moment that
$G$ is simple.  Then such a current-combination $\tilde{J}$ which meets this
condition is:
\beqn
       \tilde{J} ~=~ \left( J_G ,~ J_{SU(3)}^{(1)}, ~J_{SU(2)}^{(1)},~
                      \exp(i\sqrt{k_Y/2}\, \phi)\right)~.
\label{combJ}
\eeqn
Here $\phi$ is free boson that realizes $U(1)_Y$ via the current
$J_Y \equiv i\sqrt{k_Y/2}\partial \phi$, normalized
in accordance with the convention
(\ref{k1def}) for a $U(1)$ algebra at level $k_Y$.
In (\ref{combJ}), the choice for $J_G$ depends on the choices for the
particular factors (\ref{alphaparts}) which enter
into the expression $\alpha_G(R)$ in (\ref{quantform}), as well as on
the particular choices for the integers $z$ (which are unconstrained).
In particular, we see that these contributions
to $\alpha_G(R)$
are successfully reproduced in the monodromy
of $\tilde{J}$ with a given state
$\psi_i$ if we choose
\beqn
      SU(n+1):&~~~~~
      &J_G ~=~ J_A^{(z)} ~~~{\rm for} ~z=0,...,n~\nonumber\\
      SO(2n+1):&~~~~~ &J_G ~=~ \cases{
           {\bf 1}   & for $z=0$\cr
           J_B   & for $z=1$\cr}\nonumber\\
      SO(2n+1):&~~~~~ &J_G ~=~ \cases{
           {\bf 1}   & for $z=0$\cr
           J_C   & for $z=1$\cr}\nonumber\\
      SO(4m+2):&~~~~~
      &J_G ~=~ \cases{
     {\bf 1}& for $z=0$\cr
     J_D^{(s)}& for $z=1$\cr
     J_D^{(v)}& for $z=2$\cr
     J_D^{(c)}& for $z=3$\cr}\nonumber\\
      SO(4m):&~~~~~
      &J_G ~=~ \cases{
     {\bf 1}& for $(z_1,z_2)=(0,0)$\cr
     J_D^{(s)}& for $(z_1,z_2)=(0,1)$\cr
     J_D^{(v)}& for $(z_1,z_2)=(1,0)$\cr
     J_D^{(c)}& for $(z_1,z_2)=(1,1)~.$\cr}
\label{jgchoices}
\eeqn
It is straightforward to generalize (\ref{jgchoices}) to
the cases when the hypercolor
group $G$ is non-simple.

Thus, making the choices (\ref{jgchoices}), we find that
the total monodromy of $\tilde{J}$
with respect to an arbitrary representation $(R,r,\tilde{r},Y)$
is precisely the left side of the charge-quantization condition
(\ref{quantform}).
Since we are requiring that this is zero modulo 1, the locality
condition on the current $\tilde{J}$ is then automatically satisfied.
Although not explicitly indicated in (\ref{combJ}), the current $\tilde{J}$
also
contains an anti-holomorphic part, and we must choose this in such a way that
it too
is a local simple current. This is easily achieved if we take this current to
be the
 {\it identity}\/ primary field in the anti-analytic sector. As this
has zero conformal dimension, the condition for modular
invariance under $T:\,\tau\to \tau +1$ becomes the condition
that the conformal dimension of $\tilde{J}$ in the analytic sector
must itself also be an integer. Thus we discover that
for a consistent theory, we must have
\beqn
       h(\tilde{J}) ~=~ h(J_G) ~+~ {k_3\over 3} ~+~
          {k_2 \over 4} ~+~ {k_Y\over 4}~\in~ \IZ~
\label{Jconfdim}
\eeqn
where $h(J_G)$ represents the
conformal dimension of the relevant current $J_G$
in (\ref{jgchoices}).
This conformal dimension can be generally calculated
in terms of the conformal dimensions of the elementary simple currents
given in  (\ref{suh}), (\ref{sooddh}),
(\ref{sph}), and (\ref{soevendim}).
Hence, writing our arbitrary hypercolor gauge group in the general form
\beq
   G~\equiv~ \left\lbrack \prod_i SU(p_i)_{k_i}\right\rbrack ~\otimes~
       \left\lbrack \prod_j  SO(2q_j +1)_{k_j} \right\rbrack~\otimes~
       \left\lbrack \prod_\ell Sp(2r_\ell)_{k_\ell}\right\rbrack ~\otimes~
       \left\lbrack \prod_m SO(2s_m)_{k_m}\right\rbrack~,
\label{hypercolorgroup}
\eeq
we immediately obtain our final general condition on the \KM\ levels:
\beq
    \sum_{i} {z k_i (p_i - z) \over 2p_i}
    + \sum_{j} {k_j\over 2}
     + \sum_{\ell} {k_\ell q_\ell\over 4}
    + \sum_{m} h_m +
     {k_3\over 3} + {k_2 \over 4} + {k_Y\over 4} ~\equiv~ 0~~~({\rm mod}~1)~.~~
\label{levelcond}
\eeq
Note that in this equation, the quantity $h_m$
should be taken to be ${k_m s_m/ 8}$ if the quantization condition
requires $J_D^{(s)}$ or $J_D^{(c)}$ in $\tilde{J}$, while we should take
$h_m = k_m/2$ if the quantization condition requires $J_D^{(v)}$.
Both cases should be considered in order to allow for all possibilities.
Likewise, the integer $z$ which appears in the first term of (\ref{levelcond})
is arbitrary within the range $1\leq z\leq p_i-1$, and so once again
all possibilities must be included.
The corresponding integers $z$ and $(z_1,z_2)$
for the $SO$ and $Sp$ groups have been omitted from (\ref{levelcond}),
since their only non-trivial allowed value is $1$.
Also note that despite the
previous distinctions between the $SO(4m)$ and $SO(4m+2)$
gauge groups, all $SO(2n)$ groups now contribute
identically to this final constraint.

Eq.~(\ref{levelcond}), then, is the general relation
between the \KM\ levels that we have been seeking.  In particular,
this equation must be
satisfied in any modular-invariant string model
if that model is to contain only those states which
will bind into color singlets under the influence of an extra arbitrary
hypercolor gauge group of the form (\ref{hypercolorgroup}).
It is thus straightforward to examine the possible cases of different
hypercolor groups in order to determine which values of $(k_Y,k_2,k_3)$ are
mutually consistent.

The first special case to consider, of course,
is that for which there is {\it no}\/ hypercolor group at all, so
that the only states which appear in the spectrum of a given string model are
integer-charged directly.  In this case, the condition (\ref{levelcond})
reduces to the condition originally obtained by Schellekens \cite{Schellekens}:
\beqn
     {k_3\over 3} ~+~ {k_2 \over 4} ~+~ {k_Y\over 4} ~\equiv~ 0
                ~~~~~~({\rm mod}~1)~.
\label{simplelevelcond}
\eeqn
It is easy to see, then, that for $k_2=k_3=1$, the minimum
allowed value of $k_Y$ is indeed $k_Y=5/3$.
Thus, for level-one string models with only integer-charged states,
we have
\beq
                      k_Y ~\geq~ 5/3~.
\eeq

In general, though, string models {\it do}\/ contain fractionally charged
states.
In fact, as shown in Ref.~\cite{Schellekens},
it is impossible to have a level-one $SU(3)\times SU(2)\times U(1)_Y$
string model with $k_Y=5/3$ {\it without}\/ having fractionally charged
states in the corresponding spectrum, for any GSO projections that would
remove all of the fractionally charged states
in such cases would also simultaneously promote
the $SU(3)\times SU(2)\times U(1)_Y$ gauge symmetry to level-one $SU(5)$.
We therefore seek to determine the more general $k_Y$ constraints that
arise from (\ref{levelcond})
when arbitrary hypercolor binding groups are present.
More specifically,
motivated by the phenomenologically preferred levels (\ref{preferredvalues}),
we seek to know for which choices of binding groups
$G$ such levels may be accommodated without giving rise to
unconfined
fractionally charged states in the string spectrum.

Our procedure is as follows.
For each value of $k_2=k_3=1$ or $2$, we
examine each possible choice of binding group $G$.
Although the formalism presented above is completely general, we
restrict ourselves in this analysis to the case of simple groups only.
For each simple group, we then examine each possible rank and level
subject to the requirements that the resulting group actually {\it confine},
and that it (along with the MSSM gauge factors) not exceed the total
central charge $c_{\rm left}=22$ for the internal worldsheet theory
of a consistent heterotic string.
It turns out that there is also a periodicity in the allowed
solutions for $k_Y$ as a function of the level $k$ of the confining
group;  for example, in the case of the $SO(2n+1)$ confining groups,
we see from (\ref{levelcond}) that the allowed solutions for $k_Y$
depend only on whether $k_{SO(2n+1)}$ is even or odd.
We can therefore restrict ourselves, without loss of
generality, to the lowest levels $k$
appropriate for each confining group.
Finally, in the case of the $SU(n)$
groups, we examine each of the cases with $z=1,...,n-1$;
we likewise examine the cases with
$J_D^{(s,c)}$ and $J_D^{(v)}$ separately in the case of the $SO(2n)$
groups.

Our results are as follows.
For the simplest case with $k_2=k_3=1$,
we find that there are only sixteen different binding scenarios which
permit solutions with $k_Y<5/3$.  These confining groups, currents,
and corresponding solutions for $k_Y$ are listed below:
\beq
\begin{tabular}{c|c|c|| c | c | c}
  ~~~~$G$ ~~~~&  ~~~~~$J_G$ ~~~~ & ~~~~$k_Y$~~~~~       &
  ~~~~$G$ ~~~~&  ~~~~~$J_G$ ~~~~ & ~~~~$k_Y$~~~~~       \\
\hline
 $ SU(4)_3    $ &   $ J_A^{(1,3)}    $ &  7/6     &
     $ SU(13)_1   $  &  $  J_A^{(3,10)}  $   &  41/39    \\
 $ SU(6)_5    $ &   $ J_A^{(1,5)}    $ &  4/3     &
     $ SU(17)_1   $  &  $  J_A^{(7,10)}  $   &  73/51$^\ast$    \\
 $ SU(7)_3    $ &   $ J_A^{(2,5)}    $ &  23/21   &
     $ SU(17)_1   $  &  $  J_A^{(8,9)}   $  &  61/51    \\
 $ SU(9)_1    $ &   $ J_A^{(4,5)}    $ &  11/9    &
     $ SU(18)_1   $  &  $  J_A^{(7,11)}  $   &  10/9     \\
 $ SU(10)_1   $  &  $  J_A^{(3,7)}   $  &  22/15$^\ast$    &
     $ SU(19)_1   $  &  $  J_A^{(6,13)}  $   &  83/57$^\ast$      \\
 $ SU(10)_2   $  &  $  J_A^{(3,7)}   $  &  19/15   &
     $ SO(10)_5   $  &  $  J_D^{(s,c)}   $   & 7/6 \\
 $ SU(11)_1   $  &  $  J_A^{(3,8)}   $  &  43/33   &
     $ SO(18)_1   $  &  $  J_D^{(s,c)}   $   & 7/6 \\
 $ SU(12)_1   $  &  $  J_A^{(3,9)}   $  &  7/6    &
     $ SO(34)_1   $  &  $  J_D^{(s,c)}   $   & 7/6 \\
\end{tabular}
\label{results1}
\eeq
We have indicated with an asterisk those solutions for which $k_Y$
lies in the phenomenologically preferred range $1.4\leq k_Y\leq 1.5$.
We see, then, that for $k_2=k_3=1$, the
only binding scenarios permitting the phenomenologically
preferred values of $k_Y$ are:
\beq
      k_2=k_3=1,~~1.4\leq k_Y\leq 1.5~:~~~~~~~
           G= SU(10)_1,~ SU(17)_1,~ SU(19)_1~.
\eeq
Of these limited choices with such large ranks,
$SU(10)_1$ will probably be the most
feasible hypercolor group to realize in a consistent string model.

For $k_2=k_3=2$, we find that more solutions with $k_Y<5/3$ are possible.
Below we list only those hypercolor groups
that permit values of $k_Y$ in the range $1.4\leq k_Y\leq 1.5$:
\beq
\begin{tabular}{c|c|c|| c | c | c}
  ~~~~$G$ ~~~~&  ~~~~~$J_G$ ~~~~ & ~~~~$k_Y$~~~~~       &
  ~~~~$G$ ~~~~&  ~~~~~$J_G$ ~~~~ & ~~~~$k_Y$~~~~~       \\
\hline
 $ SU(4)_3    $ &   $ J_A^{(1,3)}    $ &  17/12     &
     $ SU(12)_1    $ &   $ J_A^{(3,9)}    $ &  17/12     \\
 $ SU(6)_5    $ &   $ J_A^{(1,5)}    $ &  3/2     &
     $ SU(17)_1    $ &   $ J_A^{(8,9)}    $ &  73/51     \\
 $ SU(9)_1    $ &   $ J_A^{(4,5)}    $ &  13/9     &
     $ SO(18)_1    $ &   $ J_D^{(s,c)}    $ &  17/12     \\
 $ SU(10)_2    $ &   $ J_A^{(3,7)}    $ &  22/15     &
     $ SO(34)_1    $ &   $ J_D^{(s,c)}    $ &  17/12     \\
 $ SU(11)_1    $ &   $ J_A^{(3,8)}    $ &  49/33     &
     $ ~  $ &   $ ~   $ &  ~    \\
\end{tabular}
\label{results2}
\eeq
Thus, for $k_2=k_3=2$, we see that
relatively small hypercolor
groups can now confine the fractionally
charged states that arise
for $1.4\leq k_Y\leq 1.5$.  In particular, $SU(4)_3$ and $SU(6)_5$
are the most likely candidates for realization in
a consistent higher-level string model.
It is straightforward to continue this analysis to higher levels $(k_2,k_3)$
and to non-simple (tensor-product) hypercolor groups.

\subsection{Constraints with Fractional Charges}

Another interesting variant of the Schellekens condition
is to suppose that the gauge groups that couple to states
with $U(1)_{\rm e.m.}$ charge are just the MSSM gauge
group factors (\ie, that $G=1$), but to
allow levels that lead
to {\it fractionally}\/ charged
asymptotic states.
One would then hope that when realized in a consistent string
model, such states can become superheavy and not appear in the
low-energy effective theory.
If we allow
arbitrary $1/p$ multiples of the electron charge, then the quantization
condition analogous to (\ref{quantform}) is
\beqn
   p\,\left({c_{su(3)}(r)\over 3} + { c_{su(2)}(\tilde{r})\over 2} + Y\right)
      ~\equiv~ 0~~~~~~~({\rm mod}~1)~.
\label{pthquant}
\eeqn
Thus, if we define the residues $p_2$ and $p_3$
via $p\equiv p_m$~mod~$m$ with $0\leq p_m<m$,
then the simple current combination
\beqn
      \tilde J ~=~ \left(J_{SU(3)}^{(p_3)},~
      J_{SU(2)}^{(p_2)}, ~\exp(ip\sqrt{k_Y/2}\,\phi)\right)
\label{pthcomb}
\eeqn
satisfies the necessary locality property.  Here $\phi$, as in (\ref{combJ}),
is a free boson realizing the $U(1)_Y$ gauge symmetry.
Computing the conformal
dimension of $\tilde J$ then gives the following constraint on
the \KM\ levels $(k_Y,k_2,k_3)$:
\beqn
      2\,p_3\, ( 3 - p_3)\,k_3 ~+~
        3\, p_2\, ( 2 - p_2)\,k_2 ~+~ 3\,p^2 \,k_Y
        ~\equiv~ 0~~~~~~({\rm mod}~12)~.
\label{ptheqn}
\eeqn

It is straightforward to tabulate the allowed values of $k_Y$
as a function of charge fractions $p$.
Clearly, as $p$ increases, the constraint (\ref{ptheqn}) becomes
weaker,
so that increasingly many solutions with $k_Y$ in any given range can
be found.
Therefore, motivated by (\ref{preferredvalues}), we seek solutions
with $k_Y$ in the narrow range $1.4\leq k_Y/k_2\leq 1.5$ for relatively
small values of $p$ and for $k_2=k_3$.

Our results are as follows.
For $k_2=k_3=1$, we find that
there are no suitable fractional charges which will allow
$1.4\leq k_Y\leq 1.5$ until $p=4$, which permits the solution
$k_Y=17/12  \approx  1.417$.
The next solution then appears for $p=6$, and has the (somewhat better)
value $k_Y=13/9\approx 1.444$.
Solutions then appear for $p=7$, and so forth.

For $k_2=k_3=2$, by contrast, we find solutions
in the desired range starting at $p=3$, which yields $k_Y/k_2=13/9$.
There are then solutions for $p=4$ and $p=5$, with
values $k_Y/k_2=17/12$ and $k_Y/k_2=107/75\approx 1.427$ respectively.
Somewhat higher values of $k_Y/k_2$ do not appear until $p=6$,
which permits both $k_Y/k_2=13/9$ and $k_Y/k_2 =3/2= 1.5$.
There are then suitable solutions for $p=7$, and so forth.
Likewise, for $k_2=k_3=3$, there are again no solutions in
the desired range until $p=4$.

Interestingly, we find that phenomenologically viable solutions
with {\it half}\/-integrally charged states ($p=2$) do
not occur until level $k_2=k_3=4$,
where we have the isolated solution $k_Y/k_2=17/12$.
Thus, models with only integrally or half-integrally charged states
cannot have values of $k_Y$ in the desired range unless their MSSM factors
are realized at level $k_2=k_3\geq 4$.
At such levels, the next solution then appears for $p=3$,
with $k_Y/k_2=13/9$.

\setcounter{footnote}{0}
\section{Conclusions and Discussion}

In this paper, we have examined the extent to which
the appearance of higher-level \KM\ gauge symmetries and non-standard
hypercharge normalizations can provide a resolution to the gauge-coupling
unification problem in string theory.
We analyzed the phenomenological constraints
on the allowed regions of $(k_Y,k_2,k_3)$ parameter space,
and found strong correlations between the values of the ratios $r\equiv
k_Y/k_2$
and $r'\equiv k_3/k_2$ and the {\it absolute sizes}\/ of the required levels
$(k_Y,k_2,k_3)$.  These results are summarized in Figs.~\ref{rvsrprime} and
\ref{rvsk2}.
We then examined
the possible hypercharge embeddings that might lead to exotic hypercharge
normalizations with $k_Y<5/3$, and found that for a
specific class of realistic string models, such values of $k_Y$ are impossible.
This class of string models includes most of the realistic string models
that exist in the literature to date.
We also considered hypercharge embeddings beyond this class, however,
and found that several isolated embeddings with $k_Y<5/3$ exist.
These embeddings involve higher-twist sectors giving rise to chiral
matter, and/or higher-level MSSM gauge group factors.
Finally, we considered the constraints on $(k_Y,k_2,k_3)$ that
arise from the requirement that no fractionally charged states
appear in the string spectrum.  We found that $k_Y\geq 5/3$ in all
string models {\it without}\/ fractionally charged states, and determined
which sets of additional hypercolor groups are capable of confining
any fractionally charged states that appear for other values of $k_Y$.
In this way we showed that only a relatively small set of hypercolor binding
scenarios are possible for the
phenomenologically preferred values of $(k_Y,k_2,k_3)$.

Our results raise a number of interesting issues
which we shall now briefly discuss.
First, as we mentioned above, it is clear that
from the results of our renormalization-group analysis
presented in Sect.~3 that there is a strong correlation
between the \KM\ ratios $r\equiv k_Y/k_2$ and the absolute
sizes of the \KM\  levels needed.
This correlation ultimately
stems from the intrinsically string-theoretic constraint that
relates the coupling at unification to the unification scale.
Thus, the \KM\ levels that are required for different values of
$k_Y/k_2$ vary markedly, and are extremely sensitive to
both experimental uncertainties (such as that in $\sin^2\theta_W(M_Z)$),
and theoretical corrections (such as those arising from heavy
string thresholds, light SUSY thresholds, and intermediate matter
thresholds).
While we have not included these latter corrections in our analysis,
their inclusion would be necessary before more precise statements
concerning the phenomenologically preferred values of $(k_Y,k_2,k_3)$
can be made.
For example, we see from the charge-quantization analysis of Sect.~6
that the appearance of non-standard values of $k_Y/k_2$ is
closely correlated to the appearance of additional fractionally-charged
states in the string spectrum.
These non-MSSM states have the potential to alter the running of the
gauge couplings significantly.  Similarly, while the effects
of the heavy string threshold corrections are typically small
in realistic string models (see, {\it e.g.}, the general arguments
presented in Ref.~\cite{unif2}), it is unclear how we can expect
these corrections to scale with the \KM\ levels $(k_2,k_3)$.
A naive analysis would suggest that the gauge-dependent parts of
these corrections should actually
scale as $k_i^{-1}$ (due to the shrinking of the roots of the charge
lattice which accompanies the appearance of higher-level gauge symmetries),
but it is possible that the subtle GSO projections that
produce these higher-level gauge symmetries can, through modular invariance,
add new twisted sectors that invalidate such naive estimates.
Likewise, if $(k_Y/k_2,k_3/k_2)$  do not take their usual MSSM values
$(5/3,1)$, then there will also be contributions from the gauge-independent
parts of these threshold corrections;  these gauge-independent pieces
are less well understood, and have the potential to be
quite large \cite{Yterm}.  Consequently, the answers to all of these
questions are highly model-dependent, and their effects cannot be included
in the sort of general framework we presented in Sect.~3.
The results of Sect.~3 should nevertheless provide a
suitable foundation upon which a realistic higher-level
string model might be constructed.

Another immediate question raised by our analysis concerns the
values of $k_Y$ that are realizable in consistent string models,
and the relation between the hypercharge embedding approach we followed
in Sects.~4 and 5, and the charge-quantization approach we
followed in Sect.~6.
In the hypercharge embedding approach, we were able to
prove that one must have $k_Y\geq 5/3$ for certain classes of string models.
By contrast, beyond this class, we found that there exist special
hypercharge embeddings which satisfy a host of string consistency
constraints and succeed in realizing hypercharge normalizations $k_Y<5/3$.
In the charge-quantization approach, we likewise found a similar situation:
for string models {\it without}\/ fractionally charged states,
we found that $k_Y\geq 5/3$, but smaller values of $k_Y$ are
potentially realizable if one allows fractionally charged states
to appear and be confined under extra non-MSSM hypercolor interactions.
Despite this superficial similarity, however, it is interesting
to note that  these two analyses did not yield identical results.
In particular, our analysis of possible hypercharge embeddings
in Sects.~4 and 5 yielded potential values of $k_Y/k_2$ which were,
in many cases, different than those found to be realizable via the
hypercolor scenarios we examined in Sect.~6.  It is therefore natural to wonder
if this observation implies that the special embeddings we constructed
in Sect.~5 are not realizable in models which are
phenomenologically acceptable ({\it i.e.}, free of observable
fractionally charged states).

Unfortunately, this is also a difficult question to answer,
for there
are various subtleties that have not been taken into account.
First,
it is clear that the charge-integrality constraints derived in Sect.~6
apply to {\it all}\/ string states, whether massive or massless,
but such constraints might be evaded if we require only that
the {\it massless}\/ states be integrally charged.  Unfortunately,
it does not seem straightforward to incorporate this additional
provision into the analysis.
Likewise, another possibility would be to impose the even weaker
requirement that only those massless states which are {\it chiral}\/ must be
integrally
charged;  after all, vector-like states which are fractionally charged and
massless at tree level can acquire potentially large masses at higher loops
({\it e.g.}, via the shift of the moduli which is generally
required in order to break anomalous $U(1)$ gauge symmetries and restore
spacetime supersymmetry \cite{DSWshift}).
Once again, however, such scenarios are highly model-dependent,
and therefore cannot be readily incorporated into the sort of general
analysis we have been conducting.

Finally, it is also important to note that
although we have restricted our analysis of $k_Y<5/3$ hypercharge embeddings
in Sects.~4 and 5 to those ``minimal'' embeddings
in which the non-abelian MSSM gauge factors
occupy only the first few components of the charge lattice,
there may also exist alternative ``extended'' embeddings
which make use of the additional lattice components \cite{shy}.
Such embeddings also have the potential to yield hypercharge
normalizations $k_Y<5/3$ together with the MSSM matter content.  However, since
such embeddings make use of the non-MSSM components of the charge lattice,
their structure is
significantly less constrained than that of the embeddings we have
studied here.
Such embeddings are therefore best analyzed on a model-by-model basis.

Thus, we conclude that there exist various
possible avenues by which
higher-level \KM\ gauge symmetries and non-standard hypercharge normalizations
can be employed to reconcile string-scale unification with the low-energy
couplings.
Indeed, given our general constraints concerning the
allowed values of $(k_Y,k_2,k_3)$,
the existence of hypercharge embeddings with $k_Y/k_2<5/3$, and
the implications of such hypercharge normalizations for the appearance
of fractionally charged string states,
it may prove possible
to systematically construct realistic string models which not only realize
the phenomenologically preferred values of $(k_Y,k_2,k_3)$, but which
also simultaneously avoid unconfined fractionally charged states.
These avenues therefore deserve further exploration.

\bigskip
\medskip
\leftline{\large\bf Acknowledgments}
\medskip

We are pleased to thank I. Antoniadis, E. Witten, and especially
S. Chaudhuri and J. Lykken for discussions.
This work was supported in part by the Department of Energy
under Grants No.\ DE-FG-0290ER40542
and  DE-FG-0586ER40272,
by the National Science Foundation under
Grant No.\ PHY94-07194, and by the W.M. Keck Foundation.


\setcounter{section}{0}   

\Appendix{MSSM Hypercharge Assignments}

In this Appendix, we briefly review the standard anomaly-cancellation
arguments which fix (uniquely, up to an overall scale) the
hypercharge assignments for the MSSM representations.

In this paper, we are considering theories that contain
the field content of the $N=1$ MSSM in their massless spectra.
This means that the spectrum of any such theory
must include the following $SU(3)\times SU(2)\times U(1)_Y$
$N=1$ superfield representations:
\beqn
             Q  &\equiv&  (3,2)_{y_Q}  \nonumber\\
             u_R  &\equiv&  (\overline{3},1)_{y_u}  \nonumber\\
             d_R  &\equiv&  (\overline{3},1)_{y_d}  \nonumber\\
             L  &\equiv&  (1,2)_{y_L}  \nonumber\\
             e_R  &\equiv&  (1,1)_{y_e}  \nonumber\\
             H^+ &\equiv&  (1,2)_{y_+}  \nonumber\\
             H^- &\equiv&  (1,2)_{y_-}  ~.
\label{representations}
\eeqn
We have left the seven hypercharge assignments
    $\lbrace y_Q,y_u,y_d,y_L,y_e,y_+,y_-\rbrace$
arbitrary, but in general
these are subject to a variety of constraints.
First, there are the constraints which arise from cancellation
of gravitational and gauge anomalies:
\beqn
      {\rm gauge/gravitational}&\Longrightarrow& {\rm Tr}\, Y~=~0\nonumber\\
      {\rm pure~gauge}&\Longrightarrow& \cases{
        {\rm Tr}\, Y^3~=~0\cr
        {\rm Tr}\, Y\,[J^{(2)}]^2~=~0\cr
        {\rm Tr}\, Y\,[J^{(3)}]^2~=~0\cr}
\label{anomalyconstraints}
\eeqn
where $J^{(2)}$ and $J^{(3)}$ refer to $SU(2)$ and $SU(3)$ currents
respectively.
Similarly, there are the constraints that come from
the required $U(1)_Y$-invariance of our desired mass terms
\beq
           Q u_R H^+~,~~~
           Q d_R H^-~,~~~
           L e_R H^-~,~~~
           H^+ H^-~,
\label{masstermconstraints}
\eeq
namely that the sum of the hypercharges for each mass term
in (\ref{masstermconstraints}) must vanish.
It turns out that only six of these eight constraint
equations are linearly independent, however.
For example, the sum of the $Q u_R H^+$
and $Q d_R H^-$ constraints reproduces the $\Tr Y [J^{(3)}]^2$ constraint
when the $H^+ H^-$ constraint is taken into account.
Taken together, therefore, these equations constrain the seven hypercharge
variables only up to an overall scale, with the fixed ratios
\beq
    y_Q:y_u:y_d:y_L:y_e:y_+:y_- ~=~ 1:-4:2:-3:6:3:-3~.
\eeq
Choosing a scale so that the singlet representation $e_R$
has unit hypercharge then yields the hypercharge assignments
\beq
    \lbrace y_Q,y_u,y_d,y_L,y_e,y_+,y_-\rbrace ~\equiv~
      \lbrace 1/6,-2/3,1/3,-1/2,1,1/2,-1/2\rbrace~.
\label{hypercharges}
\eeq
These are of course the standard hypercharge assignments,
and our point here has been to demonstrate that except for an overall
scale, there is no freedom to adjust these assignments.
Furthermore, fixing the scale by assigning unit hypercharge
to the singlet representation is convenient since this assignment
is independent of any normalizations, conventions, or \KM\ levels for the
$SU(2)$ or $SU(3)$ gauge group factors in the model.
This choice for the hypercharge scale is also the one that leads to the
usual normalization factor $k_Y=5/3$ for an $SU(5)$ or $SO(10)$ embedding.
Thus, any consistent and realistic string model containing
the MSSM spectrum  (\ref{representations}) and gauge group
$SU(3)_C\times SU(2)_L \times U(1)_Y$
must have an identifiable $U(1)_Y$ factor whose eigenvalues
acting on these representations
reproduces (\ref{hypercharges}).

\Appendix{Level-Dependence of Two-Loop Corrections}

In this Appendix we demonstrate that, as claimed in Sect.~3,
the two-loop corrections $\Delta_Y^{(\rm 2-loop)}$, $\Delta_2^{(\rm 2-loop)}$,
and $\Delta_3^{(\rm 2-loop)}$ depend only logarithmically
on the ratios of the \KM\ levels $k_Y$, $k_2$, and $k_3$.

We begin by considering the full two-loop Yukawa-less renormalization
group equations for the MSSM couplings
\beq
       {dg_i\over dt} ~=~ -{g_i\over 16\pi^2} \,\left\lbrack
          b_i g_i^2 ~+~ {1\over 16\pi^2}\,\sum_j\, b_{ij}\,g_i^2
\,g_j^2\right\rbrack~,
\eeq
which can be rewritten in terms of $\alpha_i\equiv g_i^2/(4\pi)$ as
\beq
      {d\over dt}\left({4\pi\over \alpha_i}\right)
           ~=~ 2\,b_i ~+~ {1\over 2\pi}\,\sum_j \,b_{ij}\,\alpha_j~.
\label{twoloop}
\eeq
Here $b_i$ and $b_{ij}$ are respectively the one- and two-loop beta-function
coefficients, and $t\equiv \ln(M_{\rm string}/M)$ where $M$ is the variable
setting the energy scale.
If we neglect the two-loop term in (\ref{twoloop}),
it is straightforward to integrate this equation from $M=\mu$ to $M=M_{\rm
string}$
to obtain the usual one-loop RGE
\beq
           {4\pi\over\alpha_i(\mu)} ~=~
           {4\pi\over\alpha_i(M_{\rm string})}   ~+~ 2\,b_i\,\ln{M_{\rm
string}\over \mu}~.
\label{oneloopresult}
\eeq
Thus the two-loop ``correction term'', obtained by taking the difference
between
the results of the one- and two-loop evolutions over this range, is simply
\beq
      \Delta_i^{(\rm 2-loop)} ~\equiv~ -{1\over 2\pi} \,\sum_j\,b_{ij}\,
\int_\mu^{M_{\rm string}}dt~\alpha_j(t)~.
\label{genexpression}
\eeq
We can approximate the value of this integral by substituting the one-loop
result
for $\alpha_j(t)$ from (\ref{oneloopresult}) into the integrand.
We then have
\beq
        \Delta_i^{(\rm 2-loop)}~=~ - {1\over 2\pi}\,\sum_j\,b_{ij}\,
\alpha_j(M_{\rm string})\,
         \int_\mu^{M_{\rm string}} dt\,\left\lbrack
           1~+~ {b_j\,\alpha_j(M_{\rm string})\over 2\pi}
\,t\right\rbrack^{-1}~,
\eeq
which can be integrated analytically to yield
\beq
        \Delta_i^{(\rm 2-loop)}~=~ \sum_j\,b_{ij}\,{b_j}^{-1} \ln\left\lbrack
           1~+~ {b_j \,\alpha_j(M_{\rm string})\over 2\pi}\,\ln {M_{\rm
string}\over \mu}\right\rbrack~.
\eeq
Thus we see that $\Delta_i^{(\rm 2-loop)}$
depends only logarithmically on the values of the $\alpha_i(M_{\rm string})$.
Now, $M_{\rm string}$ is defined as that scale at which the unification
takes place, with each coupling related to the string coupling at tree-level
via
\beq
        \alpha_i(M_{\rm string})~=~ {\alpha_{\rm string}(M_{\rm string})\over
k_i}~.
\eeq
Thus we conclude that the dependence of the $\Delta_i^{(\rm 2-loop)}$ on the
levels $k_i$
is also logarithmic:
\beq
        \Delta_i^{(\rm 2-loop)}~=~ \sum_j\,b_{ij}\,{b_j}^{-1} \ln\left\lbrack
           1~+~ {b_j \,\alpha_{\rm string}(M_{\rm string})\over 2\pi\,k_j}\,
        \ln {M_{\rm string}\over \mu}\right\rbrack~.
\label{kdependence}
\eeq
In fact, it is straightforward to see from (\ref{kdependence}) that this
logarithmic dependence involves only the {\it ratios}\/ of the \KM\ levels, for
if we take $\mu=M_Z$ and solve (\ref{neededlater}) for $\alpha_{\rm
string}(M_{\rm string})$
in terms of the fixed low-energy electromagnetic coupling $a\equiv \alpha_{\rm
e.m.}(M_Z)$,
we find
\beq
   \Delta_i^{(\rm 2-loop)}~=~\sum_{j} \,b_{ij} \,b_j^{-1}\, \ln\left\lbrack
   1~+~{k_Y + k_2 \over k_j}\,\left(
   {2\pi \over a\,b_j\,\ln(M_{\rm string}/M_Z)} -
   {b_Y + b_2\over b_j}\right)^{-1}
   \right\rbrack~.
\label{twoloopthresh}
\eeq
Thus, neglecting the level-dependence of the scale $M_{\rm string}$
(which appears only as a double-log effect),
we see that the leading dependence of the corrections $\Delta_i^{(\rm 2-loop)}$
on the levels $k_i$ is only through the {\it logarithm of their ratios}.
Moreover,
explicitly evaluating (\ref{twoloopthresh}) for the MSSM
spectrum and beta-function coefficients
yields results which are close to the exact results quoted
in (\ref{twoloopcorrections}), thereby verifying the validity
of the approximation
made by substituting the one-loop result (\ref{oneloopresult})
into the integrand of (\ref{genexpression}).


\vfill\eject

\bibliographystyle{unsrt}

\end{document}